\documentclass[aps,prx,reprint,twocolumn,10pt,superscriptaddress,floatfix]{revtex4-2}

\usepackage[utf8]{inputenc}
\usepackage[T1]{fontenc}
\usepackage{graphicx}
\usepackage[
  protrusion=true,
  expansion=false
]{microtype}
\usepackage{amsmath,amssymb}
\usepackage{siunitx}
\DeclareSIUnit{\gauss}{G}
\usepackage{hyperref}
\usepackage{booktabs}

\usepackage[caption=false]{subfig}
\newcommand{\phantomsubfloat}[1]{
    {
        \captionsetup[subfigure]{labelformat=empty}
        \subfloat[][]{#1}
    }%
}

\newcommand{\captionlabel}[1]{(\protect\subref{#1})}

\usepackage{bm}

\usepackage[dvipsnames]{xcolor}

\definecolor{MyBlue}{rgb}{0.15,0.3,0.8}
\definecolor{MyRed}{rgb}{0.8,0.0,0.}
\definecolor{MyGreen}{rgb}{0.13,0.55,0.13}

\definecolor{RedQ}{cmyk}{0.04,0.87,0.89,0.25}
\definecolor{RedRaman}{cmyk}{0.04,0.87,0.89,0.1}
\definecolor{RedBragg}{cmyk}{0.04,0.87,0.89,0.4}

\newcommand{\Veff}{V_{\text{eff}}}

\newcommand{\hc}{\text{h.c.}}

\newcommand{\Erec}{E_{\text{rec}}}
\newcommand{\vrec}{v_{\text{rec}}}

\newcommand{\sigmabold}{\ensuremath{\bm{\sigma}}}

\providecommand{\E}{\ensuremath{\mathrm{e}}}
\providecommand{\I}{\ensuremath{\mathrm{i}}}

\providecommand{\ket}[1]{\ensuremath{\left|#1\right\rangle}}

\providecommand{\ebold}{\ensuremath{\bm{e}}}

\providecommand{\hbold}{\ensuremath{\bm{h}}}

\providecommand{\qbold}{\ensuremath{\bm{q}}}

\providecommand{\Abold}{\ensuremath{\bm{A}}}

\newcommand\fig[1]{Fig.~\ref{fig:#1}}

\makeatletter
\def\section{\@startsection{section}{1}{\z@}{3.5ex plus 0.5ex minus 0.2ex}%
{1ex plus 0.2ex}{\normalfont\normalsize\bfseries\centering}}
\def\subsection{\@startsection{subsection}{2}{\z@}{2.5ex plus 0.4ex minus 0.2ex}%
{0.7ex plus 0.2ex}{\normalfont\normalsize\bfseries\centering}}
\def\subsubsection{\@startsection{subsubsection}{3}{\z@}{1.5ex plus 0.3ex minus 0.2ex}%
{0.6ex plus 0.2ex}{\normalfont\normalsize\itshape\centering}}
\makeatother

\hypersetup{colorlinks=true,linkcolor=black,citecolor=blue,urlcolor=blue}

\begin{document}

\title{A two-dimensional realization of the parity anomaly}

\author{Nehal Mittal}
\affiliation{Laboratoire Kastler Brossel, Collège de France, CNRS, ENS-PSL University,
Sorbonne Université, 11 Place Marcelin Berthelot, 75005 Paris, France}
\author{Tristan Villain}
\affiliation{Laboratoire Kastler Brossel, Collège de France, CNRS, ENS-PSL University,
Sorbonne Université, 11 Place Marcelin Berthelot, 75005 Paris, France}
\author{Mathis Demouchy}
\affiliation{Laboratoire Kastler Brossel, Collège de France, CNRS, ENS-PSL University,
Sorbonne Université, 11 Place Marcelin Berthelot, 75005 Paris, France}
\author{Quentin Redon}
\affiliation{Laboratoire Kastler Brossel, Collège de France, CNRS, ENS-PSL University,
Sorbonne Université, 11 Place Marcelin Berthelot, 75005 Paris, France}
\author{Raphael Lopes}
\affiliation{Laboratoire Kastler Brossel, Collège de France, CNRS, ENS-PSL University,
Sorbonne Université, 11 Place Marcelin Berthelot, 75005 Paris, France}
\author{Youssef Aziz Alaoui}
\affiliation{Laboratoire Kastler Brossel, Collège de France, CNRS, ENS-PSL University,
Sorbonne Université, 11 Place Marcelin Berthelot, 75005 Paris, France}
\author{Sylvain Nascimbene}
\email{sylvain.nascimbene@lkb.ens.fr}
\affiliation{Laboratoire Kastler Brossel, Collège de France, CNRS, ENS-PSL University,
Sorbonne Université, 11 Place Marcelin Berthelot, 75005 Paris, France}

\begin{abstract}
Quantum anomalies arise when symmetries of a classical theory cannot be preserved upon quantization, leading to unconventional topological responses. A prominent example is the parity anomaly of a single two-dimensional Dirac fermion, which enforces a half-quantized Hall response. Anomaly inflow mechanism allows this effect to be observed at the surfaces of three-dimensional topological insulators, however, its realization in a genuinely two-dimensional system has remained elusive.
Here we report the observation of a parity-anomalous Hall response at the critical point of a quantum Hall topological phase transition in a synthetic two-dimensional system of ultracold dysprosium atoms. By coupling a continuous spatial dimension to a finite synthetic dimension encoded in atomic spin states, we engineer tunable Chern bands with $\mathcal{C}=0$ and $1$. At the transition, the bulk gap closes at a single Dirac point, where we observe a robust half-quantized Hall drift despite strong non-adiabatic excitations. We show that this response originates from the global structure of the band topology, is protected by an emergent parity symmetry at criticality, and disappears when parity is explicitly broken. Our work establishes synthetic quantum systems as a powerful platform to probe quantum anomalies and their interplay with topology and non-equilibrium dynamics.
\end{abstract}

\maketitle

\section{Introduction}

Symmetries play a central role in physics, constraining dynamics and protecting conserved quantities. Yet, in quantum field theory, certain classical symmetries cannot be preserved upon quantization, due to the incompatibility between symmetry requirements and the regularization needed to define quantum fluctuations. Though originally studied in particle physics \cite{bertlmann_anomalies_2000}, quantum anomalies also underlie robust phenomena in condensed matter systems \cite{fradkin_field_2013}. A paradigmatic example is the chiral (Adler-Bell-Jackiw) anomaly, involved in the decay of the neutral pion $\pi^0$ \cite{adler_axial-vector_1969,bell_pcac_1969}, and also realized in Dirac-Weyl semimetals as a negative longitudinal magnetoresistance under parallel electric and magnetic fields \cite{nielsen_adler-bell-jackiw_1983,xiong_evidence_2015,huang_observation_2015,zhang_signatures_2016,ong_experimental_2021}.

Quantum anomalies can manifest in both two- and three-dimensional systems. In the integer quantum Hall effect, the quantized Hall response of the bulk is described by a topological field theory that would be inconsistent at a boundary if taken alone. This inconsistency is resolved by chiral edge modes, whose anomalous transport exactly compensates the bulk contribution. This mechanism, known as anomaly inflow, provides a field-theory understanding of bulk-edge correspondence in 2D quantum Hall systems \cite{callan_anomalies_1985,qi_topological_2008}.

Some anomalies, however, are intrinsic to two-dimensional (2D) systems and do not require a higher-dimensional bulk. A notable example is the parity anomaly of a single 2D Dirac fermion, which arises from the impossibility of simultaneously preserving gauge invariance and parity upon quantization~\cite{niemi_axial-anomaly-induced_1983,redlich_parity_1984}. In a seminal work, Haldane introduced a minimal model in which a single Dirac point emerges at a topological phase transition, with the parity anomaly giving rise to a half-quantized Hall conductance \cite{haldane_model_1988}. However, this effect is typically obscured in lattice systems by the Nielsen-Ninomiya theorem, which states that Dirac cones must appear in pairs~\cite{nielsen_no-go_1981}, preventing the observation of half-quantized responses in materials such as graphene~\cite{novoselov_two-dimensional_2005}, conventional 2D materials~\cite{yu_quantized_2010,chang_experimental_2013,serlin_intrinsic_2020,deng_quantum_2020}, and engineered photonic or atomic lattices~\cite{wang_observation_2009,jotzu_experimental_2014,klembt_exciton-polariton_2018}.

An indirect realization of the parity anomaly occurs at the surfaces of 3D topological insulators, which host an odd number of Dirac cones~\cite{fu_topological_2007,qi_topological_2008}. Magnetic gapping of these surfaces yields half-quantized surface conductance~\cite{lu_half-magnetic_2021,mogi_experimental_2022}, consistent with the parity anomaly. Critically, this surface response maintains gauge invariance through anomaly inflow: the fractional surface contribution is compensated by bulk topological transport~\cite{qi_topological_2008}.

Observing the parity anomaly in a genuinely 2D system requires accessing a critical point where a single Dirac cone emerges~\cite{haldane_model_1988}—a regime recently realized in engineered photonic and atomic platforms~\cite{leykam_anomalous_2016,liu_observation_2020,zhong_topological_2024,yuan_observation_2025}. However, measuring the half-quantized Hall response has remained elusive, hindered by the gapless nature of critical points, where non-adiabatic excitations typically obscure quantized signatures.

Here, we experimentally observe the parity anomaly at a topological phase transition in a synthetic 2D quantum system of ultracold dysprosium atoms, realizing the minimal scenario envisioned by Haldane for a Chern-insulator transition governed by a single Dirac cone. By encoding a synthetic dimension in the spin $J=8$ of this highly magnetic atom~\cite{celi_synthetic_2014,mancini_observation_2015,stuhl_visualizing_2015,chalopin_probing_2020} and coupling it to a spatial dimension, we engineer effective 2D bands with tunable Chern numbers and  a clear distinction between edge and bulk physics. By probing the local Hall response and tracking its adiabaticity, we resolve a sharp topological phase transition between a Chern and a trivial insulator, observe the delocalization of chiral edge modes into the bulk, and locate a single Dirac point at criticality. 

At the critical point, we measure strong non-adiabatic excitations develop near the Dirac node, yet the Hall response remains robustly half-quantized. This striking behavior reflects a contribution from the entire Brillouin zone and constitutes direct evidence of its anomalous origin, providing a textbook realization of the minimal Dirac-cone scenario proposed by Haldane. Our work further demonstrates how synthetic quantum systems enable controlled studies of quantum anomalies and opens new perspectives on their interplay with interactions and non-equilibrium dynamics in topological matter.

\section{Parity anomaly\\and coupled wire model}
Before describing our experiments, we introduce the concept of the parity anomaly and present an idealized model that captures the essential physics of our system.

\subsection{Parity anomaly}
We begin with the continuum Dirac Hamiltonian in two dimensions, describing low-energy excitations near a single Dirac point:
\begin{equation}
H(\mathbf{q}) = v (q_x \sigma_x + q_y \sigma_y) ,
\end{equation}
where $\mathbf{q}$ is the quasi-momentum measured from the Dirac point, and $v$ is the velocity. This Hamiltonian describes a gapless Dirac cone with linear dispersion. It exhibits a parity symmetry
\[
\mathcal P = \sigma_y , \qquad 
\mathcal P H(q_x,q_y)\mathcal P^{-1} = H(-q_x,q_y),
\]
involving a mirror symmetry across the $y$-axis combined with a $\pi$ spin rotation around $y$, and is also time-reversal invariant. For any momentum $\qbold\neq\boldsymbol0$, the ground and excited bands are separated by a finite gap, and the Berry curvature vanishes $F(\qbold\neq\boldsymbol0)=0$. The Dirac point at $\mathbf q=\boldsymbol0$ constitutes a singular point where the Bloch states are ill-defined, carrying a $\pi$ Berry phase. Because of time-reversal symmetry, it is clear that any Hall response must vanish in the gapless Dirac system, even in the presence of diabatic excitations close to the Dirac point.

An energy gap can be induced by a non-zero mass term $\Delta\, \sigma_z$ with $\Delta\neq0$, leading to a breaking of parity and time-reversal symmetries. 
For such a gapped Dirac cone, the Berry curvature of the lower band is given by
\begin{equation}
F(\mathbf q) = \frac{v^2 \Delta}{2(v^2 q^2 + \Delta^2)^{3/2}},
\end{equation}
which integrates over momentum space to produce a half-integer Hall conductance for an isolated Dirac cone:
\begin{equation}
\sigma_H = \frac{1}{2} \, \mathrm{sgn}(\Delta) \frac{e^2}{h}.
\end{equation}
This half-quantized Hall response is the hallmark of the parity anomaly in the continuum theory.

While the continuum Hamiltonian describes an isolated linear band touching, any microscopic realization requires regularization over the full Brillouin zone. A minimal lattice regularization replaces the linear dispersion with periodic functions of quasi-momentum, for example:
\begin{equation}
H(\mathbf q) = v(\sin q_x \, \sigma_x + \sin q_y \, \sigma_y).
\end{equation}
Parity symmetry is preserved in this regularized Hamiltonian, but the Dirac point at $\mathbf q=\boldsymbol0$ is accompanied by additional Dirac points at other high-symmetry momenta of the Brillouin zone $\mathbf q=(\pi,0)$, $(0,\pi)$, and $(\pi,\pi)$. 
This reflects  the Nielsen-Ninomiya theorem, which states that 
Dirac points must appear in pairs of opposite chirality in any lattice model that is local, translationally invariant, and Hermitian~\cite{nielsen_no-go_1981}.
Consequently, opening the gap at all Dirac points, for instance using a mass term $\Delta\sigma_z$, yields a quantized Hall conductance $\sigma_H = \mathcal{C} e^2/h$, where the Chern number $\mathcal{C}$ is an integer determined by the contributions of the even number of Dirac points.

To access the physics of a single Dirac cone and the associated half-quantized Hall response, one can introduce a more complex lattice Hamiltonian with a momentum-dependent mass term, often called a Wilson term:
\begin{align*}
H(\mathbf q) = &~v(\sin q_x \, \sigma_x + \sin q_y \, \sigma_y)  + t(2 - \cos q_x - \cos q_y)\sigma_z.
\end{align*}
Here, the $t$ term gaps out the extra Dirac points at high-symmetry momenta, leaving a single low-energy Dirac cone at $\mathbf q = 0$. In this model, parity symmetry is broken globally by the momentum-dependent mass term, but it is approximately restored near $\mathbf q=0$. The Hall conductance can be calculated by summing the contributions of the odd number of opened Dirac points, producing a half-quantized Hall conductance. The remaining Dirac gap closing at $\qbold=\boldsymbol0$ leads to a breaking of adiabaticity of any Hall response measurement, but it does not affect the overall half-quantization. This scenario was first proposed by Haldane in his seminal work on the quantum Hall effect without Landau levels~\cite{haldane_model_1988}, and has since served as a paradigmatic example of how quantum anomalies can emerge in condensed matter systems.

\subsection{Coupled wire model}
\begin{figure*}[t]
  \begin{center}
  \includegraphics[scale=0.78,trim=10 0 0 0]{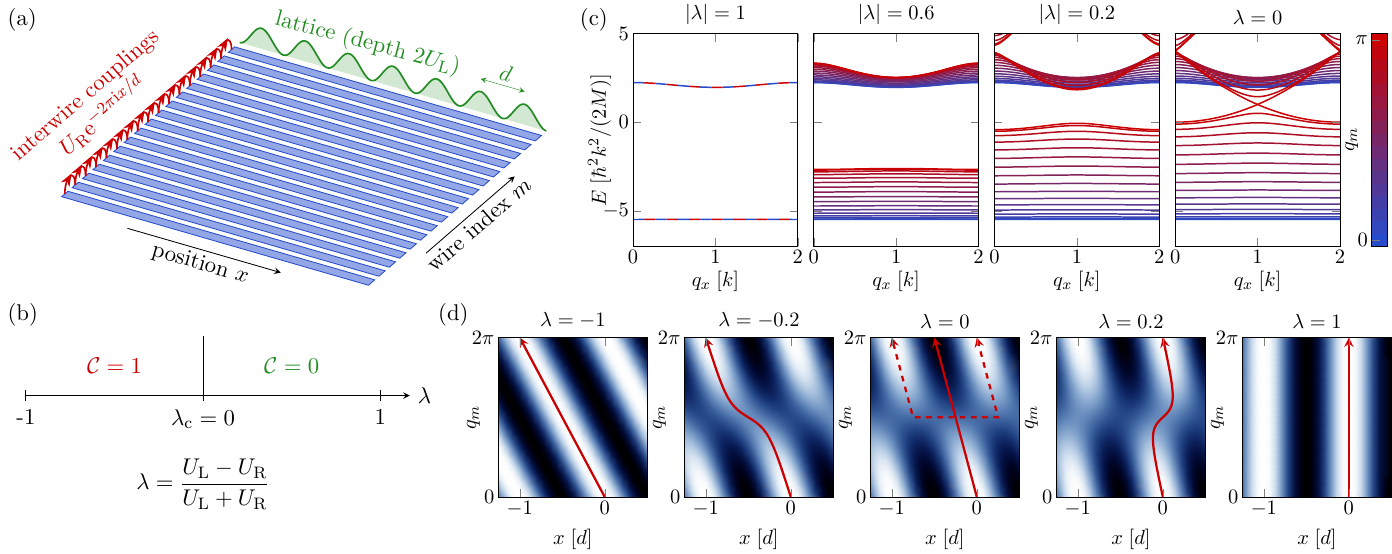}
  \end{center}
  \phantomsubfloat{\label{fig:coupled_wire_model:dimensions}}
  \phantomsubfloat{\label{fig:coupled_wire_model:phase_diagram}}
  \phantomsubfloat{\label{fig:coupled_wire_model:bandstructure}}
  \phantomsubfloat{\label{fig:coupled_wire_model:Wannier_function_drift}}
  \vspace{-3\baselineskip}
  \caption{
  \captionlabel{fig:coupled_wire_model:dimensions}
  Synthetic two-dimensional quantum system composed of a continuous spatial dimension $x$ and a discrete synthetic dimension $m$. The kinetic energy and cosine lattice potential govern dynamics along $x$, while position-dependent hoppings along $m$ acquire phases from an effective vector potential.
  \captionlabel{fig:coupled_wire_model:phase_diagram}
  Phase diagram showing tunable Chern numbers $\mathcal{C}=0$ and $1$ in gapped phases, separated by a critical point where the bulk gap closes at a single Dirac point.
  \captionlabel{fig:coupled_wire_model:bandstructure}
  Band structure of the coupled-wire model for various values of $|\lambda|$ at coupling strength $U=4.8\,\hbar^2k^2/(2M)$. Color encodes quasi-momentum $q_m$ along the synthetic dimension. At the critical point $\lambda=0$, the bulk gap closes at a single Dirac point at the $M$ point $(q_x,q_m)=(k,\pi)$.
  \captionlabel{fig:coupled_wire_model:Wannier_function_drift}
  Quantized Wannier function center drift across the phase transition. For each $q_m$, the effective potential is a cosine function whose minimum defines the Wannier function center. During an adiabatic evolution of $q_m$ from $0$ to $2\pi$, this minimum drifts by a quantized distance $-\mathcal{C}d$. At the critical point $\lambda=0$, the potential vanishes at $q_m=\pi$, leading to a discontinuity in the Wannier function center (dashed lines), reflecting a breaking of adiabaticity. However, the line $x=- q_md/\pi$ (solid line) remains a mirror symmetry axis, constraining the Wannier center to follow this line and yielding half-quantized drift despite non-adiabatic effects.
  }
  \label{fig:coupled_wire_model}
\end{figure*}

We now consider a more realistic model that captures the essential physics of our experimental system, which consists of a continuous spatial dimension $x$ coupled to a discrete dimension $m$, known as the coupled-wire model~\cite{kane_fractional_2002} (see \fig{coupled_wire_model:dimensions}). Dynamics along $x$ are governed by kinetic energy $p_x^2/2M$, where $p_x$ is momentum and $M$ is mass. Hopping between neighboring sites along the synthetic dimension, $m\rightarrow m+1$, is mediated by spatially modulated hopping amplitudes $-U_{\text{R}}\,\E^{-2\I k x}$. This spatial modulation acts like a Peierls phase from an effective vector potential $\Abold=-2\hbar k x\,\ebold_m$ coupled to particles with unit charge, generating an effective magnetic field perpendicular to the $xm$ plane with one flux quantum $\Phi_0=h$ per plaquette of area $l_x \times l_m = (d \times 1)$, where $d=\pi/k$. As shown in Ref.~\cite{kane_fractional_2002}, this coupled-wire geometry yields band structures with topological properties analogous to Landau levels. Each band carries a quantized Chern number $\mathcal{C}=1$, which determines the quantized Hall conductance~\cite{thouless_quantization_1983}.

In order to further control the bandstructure topology, the system is subjected to a lattice potential of depth $2U_{\text{L}}$ along $x$ with spacing $d$ (see \cite{szumniak_chiral_2016} for a related model). The Hamiltonian reads
\begin{align*}
H=\frac{p_x^2}{2M}-\frac{U}{2}\big[&(1-\lambda)(T_+\E^{-\I2kx}+\hc)\\
&+2(1+\lambda)\cos(2kx)\big],
\end{align*}
where $U=U_{\text{L}}+U_{\text{R}}$ is the coupling strength and $\lambda=(U_{\text{L}}-U_{\text{R}})/(U_{\text{L}}+U_{\text{R}})$ is the tuning parameter. The operator $T_+$ increments the spin projection $m$ by one unit.

Due to discrete translational invariance along both $x$ and $m$, eigenstates are labeled by quasi-momenta $q_x$ and $q_m$, defined modulo $2\hbar k$ and $2\pi$. For a given momentum $q_m$, the translation operator along $m$ can be written as $T_+=\E^{-\I q_m}$, and the inter-wire and lattice couplings at fixed $q_m$ become
\begin{align}
\Veff(q_m,x)=-U\big[&(1-\lambda)\cos(2kx+q_m)\nonumber\\
&+(1+\lambda)\cos(2kx)\big],\label{eq:Veff}
\end{align}
which simplifies to
\[
\Veff(q_m,x)=-U_{\text{eff}}(q_m)\cos[2kx+\phi(q_m)],
\]
where
\begin{align}
U_{\text{eff}}(q_m)&=U\sqrt{2[(1+\lambda^2)+(1-\lambda^2)\cos(q_m)]},\\
\tan[\phi(q_m)]&=\frac{(1-\lambda)\sin(q_m)}{(1+\lambda)+(1-\lambda)\cos(q_m)}.
\end{align}

The band structure is obtained by solving the one-dimensional Schrödinger equation with $\Veff(q_m,x)$ at each $q_m$. The ground and first excited bands are separated by a gap as long as $U_{\text{eff}}(q_m)\neq0$ for all $q_m$ (see \fig{coupled_wire_model:bandstructure}). This holds whenever $\lambda\neq0$. At the critical point $\lambda=0$, the potential vanishes at $q_m=\pi$, causing the gap to close. Since $U_{\text{eff}}(q_m)$ is an even function of $\lambda$, band structures for $\pm\lambda$ are identical.

\subsection{Topological phase diagram}
Although systems with opposite tuning parameters $\pm\lambda$ exhibit identical thermodynamic properties, their bands have distinct topological properties, revealed by analyzing the adiabatic evolution of the Wannier function center $x_c(q_m)$ as a function of $q_m$~\cite{thouless_quantization_1983,resta_theory_2007}. For a cosine lattice, this center matches the potential minimum at $x_c(q_m)=-\phi(q_m)d/(2\pi)$. As $q_m$ varies from $0$ to $2\pi$, the phase $\phi(q_m)$ winds an integer number of times $\mathcal{C}$, and $x_c$ drifts by a quantized distance $-\mathcal{C}d$. The winding number $\mathcal{C}$ then quantifies the Hall drift, corresponding to the Chern number of the ground band. 

Figure~\ref{fig:coupled_wire_model:Wannier_function_drift} illustrates the potential $\Veff(q_m,x)$ for different $\lambda$ values. For $\lambda<0$ (where $U_{\text{L}}<U_{\text{R}}$), the Wannier center exhibits a nontrivial drift of $-d$, consistent with $\mathcal{C}=1$. For $\lambda>0$, the center returns to its initial position after one period, consistent with $\mathcal{C}=0$. The value $\lambda = 0$ thus represents a critical point of the topological phase transition, where the Chern number is ill-defined due to the gap closing, which prevents an adiabatic Hall response.

\subsection{Half-quantized Hall response at criticality}

Despite the gap closing at $\lambda=0$, one still expects the Wannier center to exhibit a robust half-quantized Hall drift due to a simple symmetry argument.
At $\lambda=0$, the effective potential becomes
\[
\Veff(q_m,x)=-2U\cos\frac{q_m}{2}\cos\!\left(2kx+\frac{q_m}{2}\right),
\]
vanishing at $q_m=\pi$ and closing the gap at $q_x=k$. Adiabatic evolution of the Wannier center along the potential minimum then breaks down (see \fig{coupled_wire_model:Wannier_function_drift}). Nevertheless, the potential preserves mirror symmetry $\mathcal{M}$ about
\[
x_c(q_m)=-\frac{d\,q_m}{4\pi}.
\]
This symmetry constrains the Wannier center to follow this extremum line, yielding a half-quantized Hall drift of $-d/2$ as $q_m$ varies from $0$ to $2\pi$.

\subsubsection{Emergent parity symmetry and Dirac point}

The half-quantization originates from the emergence of a local parity symmetry that enforces a Dirac-type gap closing at criticality. To see this, we derive a low-energy effective theory near the gap-closing point $M$. 

Near $\lambda=0$ and for quasi-momenta close to the $M$ point, the inter-wire and lattice potentials nearly cancel each other, suppressing the effective potential. By projecting the dynamics onto the two lowest-energy momentum states $p_x=\hbar\left(\delta q_x \pm k\right)$, we define an effective spin-1/2 basis $\ket{\pm}$. Here, we define the quasi-momentum deviation from the $M$ point $\delta\qbold\equiv(q_x-k,q_m-\pi)$.
Restricting to this two-band subspace, the mirror symmetry $\mathcal{M}$ acts as
\[
\mathcal{M}_{\text{2-band}}\begin{pmatrix} \ket{+, \delta q_x} \\ \ket{-, \delta q_x} \end{pmatrix}
=\begin{pmatrix} 0 & e^{i \Phi} \\[2mm] e^{-i \Phi} & 0 \end{pmatrix}
\begin{pmatrix} \ket{+, -\delta q_x} \\ \ket{-, -\delta q_x} \end{pmatrix},
\]
with the momentum-dependent phase
\[
\Phi=\left(\pi+\frac{\delta q_x}{k}\right)\left(\frac{1}{2}+\frac{\delta q_m}{2\pi}\right).
\]
Near the $M$ point, $\Phi \simeq \pi/2 + O(\delta\qbold)$, so $\mathcal{M}_{\text{2-band}}$ reduces to a parity transformation combining quasi-momentum reversal $\delta q_x \to -\delta q_x$ with a $\pi$-rotation about the $y$-axis. 

This emergent parity symmetry gives rise to a Dirac-type gap closing at criticality. More precisely, we write the two-band Hamiltonian close to criticality as 
\[
H_{\text{2-band}}=\frac{\hbar^2(k^2+\delta q_x^2)}{2M}+\hbold(\delta\qbold)\cdot\sigmabold,
\]
where $\sigmabold$ denotes Pauli matrices and the effective magnetic field reads
\begin{equation*}
\hbold(\delta\qbold)=\begin{pmatrix} 
  -\frac{U}{2}[(1-\lambda)\cos q_m+(1+\lambda)]\\
  -\frac{U}{2}(1-\lambda)\sin q_m\\  
  \frac{\hbar k}{M} \delta q_x
\end{pmatrix}.
\end{equation*}
Expanding to first order in $\delta\qbold$ and $\lambda$ yields the Dirac Hamiltonian (up to an irrelevant energy offset)
\[  
H_{\text{Dirac}}=v_x\delta q_x\,\sigma_z+v_m\delta q_m\,\sigma_y -\Delta\,\sigma_x,
\]
describing a two-dimensional Dirac fermion with mass term $\Delta=U\lambda$ and velocities $v_x=\hbar k/M$ and $v_m=U/2$. The sign change of the mass across $\lambda=0$ indicates a topological phase transition with different Chern numbers~\cite{haldane_model_1988}. Figure~\ref{fig:coupled_wire_model:bandstructure} confirms the existence of a single Dirac point at criticality.

\subsubsection{The parity anomaly\label{sec:parity_anomaly}}

The emergent parity symmetry is local to the Dirac point and cannot extend globally across the Brillouin zone.  This incompatibility—preserving gauge invariance while maintaining global parity in a system with a single Dirac fermion—constitutes the parity anomaly~\cite{niemi_axial-anomaly-induced_1983,redlich_parity_1984}.

\begin{figure}[t]
  \begin{center}
  \includegraphics[scale=0.9]{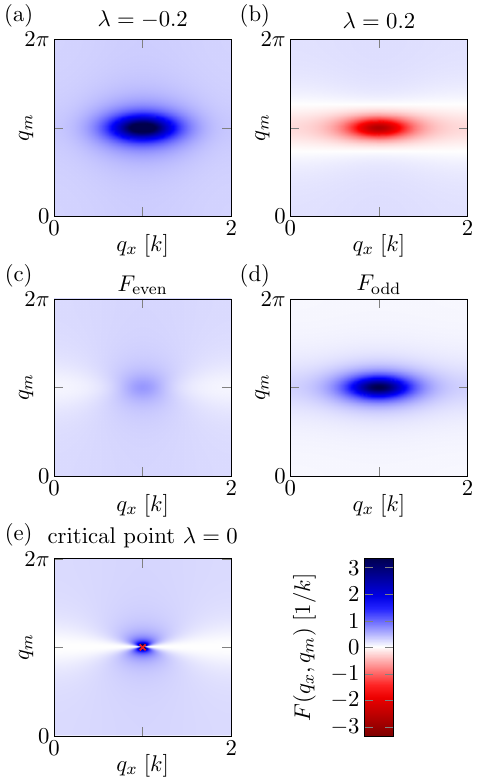}
  \end{center}
  \phantomsubfloat{\label{fig:Berry_curvature:neg}}
  \phantomsubfloat{\label{fig:Berry_curvature:pos}}
  \phantomsubfloat{\label{fig:Berry_curvature:even}}
  \phantomsubfloat{\label{fig:Berry_curvature:odd}}
  \phantomsubfloat{\label{fig:Berry_curvature:critical}}
  \vspace{-3\baselineskip}
  \caption{
    Berry curvature in the critical regime and at criticality. In the critical regime, the Berry curvature profiles at $\lambda=-0.2$ (a) and $0.2$ (b) are combined to form the even and odd contributions $F_{\rm even}(q_x,q_m)$ and $F_{\rm odd}(q_x,q_m)$ (c and d, respectively). The even contribution extends over the entire Brillouin zone, while the odd one is localized close to the $M$ point. (e) Berry curvature at the critical point, similar to the even contribution in (c). Its integral over the Brillouin zone yields a half-quantized Chern number $\mathcal{C}=0.5$. The $M$ Dirac point is shown as a red cross.
  }
  \label{fig:Berry_curvature}
\end{figure}

Beyond symmetry breaking, the parity anomaly manifests as a half-quantized Hall conductance at criticality~\cite{haldane_model_1988,semenoff_condensed-matter_1984,fu_quantum_2022}. To understand this quantitatively, we decompose the Berry curvature $F(q_x,q_m)$ of the ground band into components with opposite parity under $\lambda\to-\lambda$ (see \fig{Berry_curvature}):
\[
F(q_x,q_m)=F_{\rm even}(q_x,q_m)+F_{\rm odd}(q_x,q_m).
\]
In the critical regime, the odd component sharply peaks at the Dirac point as \cite{bernevig_topological_2013}
\[
F_{\rm odd}(q_x,q_m)\simeq-\frac{1}{2}\frac{v_x v_m}{\Delta^2}\frac{\mathrm{sgn}(\Delta)}{[1 + (v_x^2\delta q_x^2 + v_m^2\delta q_m^2)/\Delta^2]^{3/2}},
\]
contributing 
 $\mathcal C_{\rm odd}=-\mathrm{sgn}(\lambda)/2$ to the Chern number (\fig{Berry_curvature:odd}).
 Conversely, the even component varies smoothly across the entire Brillouin zone and yields $\mathcal C_{\rm even}=+1/2$  (\fig{Berry_curvature:even}). In gapped phases, these contributions sum to integer Chern numbers $\mathcal{C}\in\{0,1\}$. At criticality, the odd contribution vanishes and the Berry curvature reduces to the even component that constitutes the parity anomaly (\fig{Berry_curvature:critical}). The Dirac point constitutes a singularity at which linear response is ill-defined.

  Consequently, the measured Hall response strongly depends on the measurement protocol and the system's proximity to criticality. For any finite $\lambda \neq 0$, sufficiently weak perturbations allow the system to adiabatically follow the ground band, enabling the sharp peak $F_{\rm odd}(q_x,q_m)$ to contribute fully and restore integer quantization. When the Hall response is instead probed with a given finite perturbation strength, the system can only adiabatically follow the ground band away from the critical point. Close to criticality, non-adiabatic transitions occurring near the Dirac point suppress the localized contribution $F_{\rm odd}(q_x,q_m)$. The Hall response is then solely given by the smooth, extended even contribution $\mathcal{C}_{\rm even}=1/2$, corresponding to a half-quantized value reflecting the parity anomaly~\cite{qi_topological_2008}.

\section{Experimental implementation}

\begin{figure}[t]
  \begin{center}
  \includegraphics[scale=0.9]{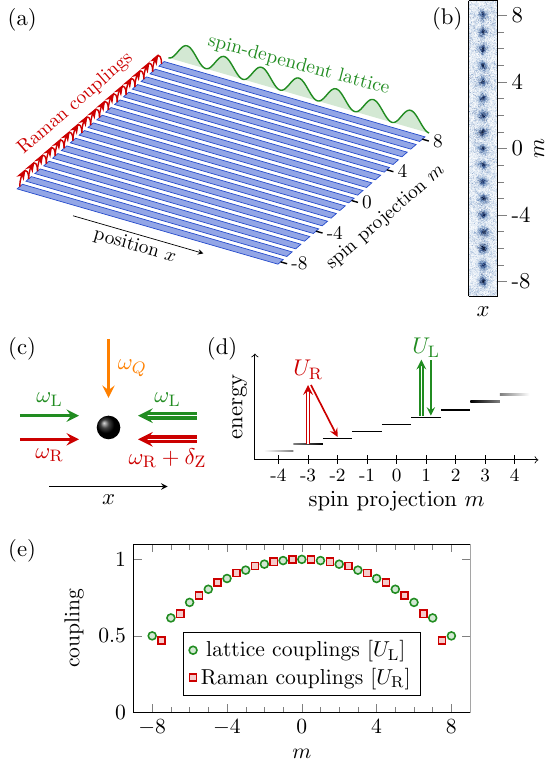}
  \end{center}
  \phantomsubfloat{\label{fig:implementation:geometry}}
  \phantomsubfloat{\label{fig:implementation:image}}
  \phantomsubfloat{\label{fig:implementation:lasers}}
  \phantomsubfloat{\label{fig:implementation:transitions}}
  \phantomsubfloat{\label{fig:implementation:couplings}}
  \vspace{-3\baselineskip}
  \caption{
    Experimental implementation of the coupled-wire model using ultracold dysprosium atoms. 
    \captionlabel{fig:implementation:geometry} Schematic of the synthetic geometry in the $xm$ plane, with sharp edges at $m=\pm J$.
    \captionlabel{fig:implementation:image} Example of an absorption image of the atomic cloud after a Stern-Gerlach magnetic pulse and a time-of-flight expansion, showing the momentum distribution along $x$ and the spin distribution along $m$ (image averaged over 40 repetitions with parameters  $\lambda=0$, $q_x=0$ and $U=4.8(2)\,\Erec$). 
    \captionlabel{fig:implementation:lasers} Two Raman beams (red arrows) counterpropagating along $x$ induce position-dependent couplings between neighboring spin states $m\rightarrow m+1$, while an optical lattice along $x$ creates a periodic potential (green arrows). An additional laser beam (orange arrow) generates a quadratic Zeeman field. \captionlabel{fig:implementation:transitions} Illustration of two-photon transitions: Raman transitions modify the spin state, while the optical lattice leaves it unchanged. 
    \captionlabel{fig:implementation:couplings} Spin-dependent coupling strengths for lattice and Raman couplings. Laser polarizations are engineered to ensure optimal matching between lattice and Raman coupling algebras.
  }
  \label{fig:implementation}
\end{figure}

\begin{figure*}
  \begin{center}
  \includegraphics[scale=0.78,trim=5 20 0 0]{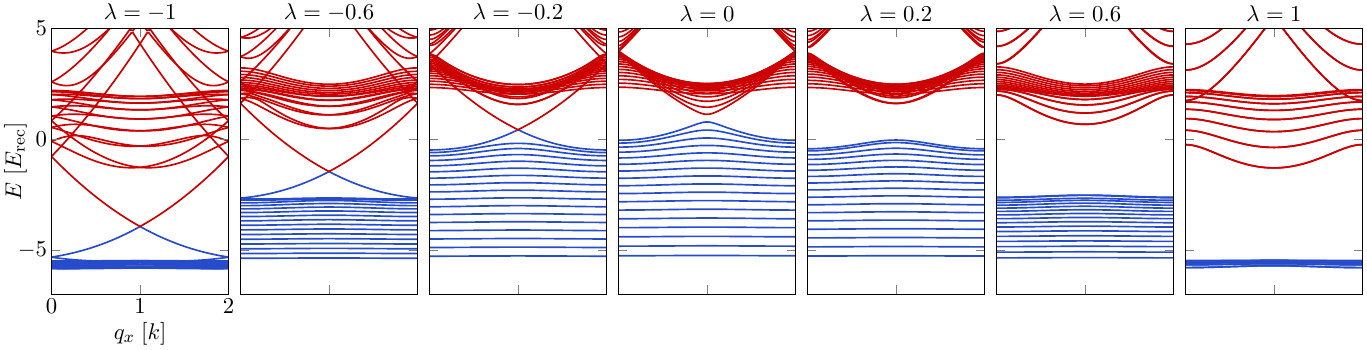}
  \end{center}
  \caption{Band structure of our implementation of the coupled-wire model for various values of the tuning parameter $\lambda$ at coupling strength $U=4.8\,\Erec$. The finite extent of the synthetic dimension causes the ground band to split into $2J+1$ subbands (blue lines). For $\lambda<0$ (topological phase), gapless chiral edge modes connect the ground and first excited bands, reflecting open boundary conditions and bulk-boundary correspondence. At criticality ($\lambda=0$), the bulk gap closes, up to finite-size effects, at a single point located at $q_x=k$.
  \label{fig:bandstructure_exp}
  }
\end{figure*}

We experimentally implement the coupled-wire model using ultracold $^{162}$Dy atoms~\cite{chalopin_probing_2020}. The atoms evolve in an effectively two-dimensional geometry with one continuous spatial dimension $x$ and one discrete synthetic dimension encoded in the $2J+1$ internal spin states, where $J=8$ is the total electronic angular momentum and $m\in[-J,J]$ is the spin projection~\cite{celi_synthetic_2014,mancini_observation_2015,stuhl_visualizing_2015,chalopin_probing_2020} (see \fig{implementation:geometry},b). 

Position-dependent couplings between neighboring spin states are generated by two counterpropagating Raman laser beams (wavelength $2\pi/k=\SI{626.1}{\nano\meter}$) along $x$  (see \fig{implementation:lasers}). These beams are detuned $\sim\SI{7}{\giga\hertz}$ from atomic resonance and frequency-shifted by the Zeeman splitting $\omega_z/2\pi=\SI{342}{\kilo\hertz}$ to resonantly couple adjacent spin states via two-photon transitions  (see \fig{implementation:transitions}). Each transition transfers momentum $2\hbar k$ along $x$, realizing the desired spatially modulated hopping. An optical lattice with the same wavelength produces a periodic potential along $x$. The light momentum defines the lattice spacing $d=\pi/k$ and the characteristic recoil energy scale $E_{\text{rec}}=\hbar^2 k^2/(2M)$, where $M$ is the atomic mass. 

A crucial experimental requirement is suppressing $m$-dependent variations in the Raman and lattice couplings. The two-photon Raman process involves rank-1 and rank-2 spin operators, which necessarily give rise to significant variations in the two-photon matrix elements across the synthetic dimension. We mitigate this by engineering the laser polarization so that Raman couplings are described by the spin-raising operator $J_+$, while the lattice depth
exhibits a weak $m$-dependent correction $U_{\rm L}(1-m^2/2J^2)$ that follows the Raman coupling inhomogeneity (Fig.~\ref{fig:implementation:couplings}). This ensures near-uniform tuning parameter $\lambda$ across all spin states, minimizing finite-size effects. The spin dependency of the light couplings also induces a dispersion of the ground band. We compensate this by applying an additional laser beam along $z$ that creates a quadratic Zeeman shift $Q m^2$ (see \fig{implementation:lasers}), which is tuned to flatten the ground band dispersion.

The resulting effective Hamiltonian is
\begin{align}      
H=\frac{p_x^2}{2M}&-\frac{U}{2}\bigg[(1-\lambda)\left(\frac{J_+}{J+1/2}\,e^{-\I2kx}+\hc\right)\nonumber\\
&-2(1+\lambda)\left(1-\frac{J_z^2}{2J^2}\right)\cos(2kx)\bigg]+ Q J_z^2,
\end{align}
where the coupling strength $U$ is controlled via laser intensity, and the tuning parameter $\lambda$ is adjusted through the relative Raman-to-lattice intensity ratio. The  band structure of the realized system (Fig.~\ref{fig:bandstructure_exp}) closely reproduces the idealized model, confirming the validity of our implementation.

Due to the finite extent of the synthetic dimension (hard-wall boundaries at $m=\pm J$), the quasi-momentum $q_m$ along the synthetic dimension--which physically represents the azimuthal phase of the spin \cite{carruthers_phase_1968}--is no longer a good quantum number. Each band $E_n(q_x,q_m)$ of the ideal infinite-system bands then splits into a set of $2J+1$ subbands $E_{n'}(q_x)$. Moreover, open boundary conditions give rise to chiral edge modes in the topological phase, manifesting bulk-boundary correspondence. At criticality, finite-size quantization yields a residual gap of order $\sim U/(2J)$ between subbands. This small value is negligible compared to the overall band dispersions and does not significantly affect the measured global band properties.

\section{Topological phase transition}
Before characterizing the parity anomaly at criticality, we investigate the topological phase transition in our system and determine the critical point.

\subsection{Hall response measurement}
We identify the topological phase diagram by measuring the Hall response across the transition as a function of the tuning parameter $\lambda$. Measuring the Chern number via Hall drift has been demonstrated in several cold atom platforms, either through real-space Hall drift~\cite{aidelsburger_measuring_2015} or circular dichroism~\cite{asteria_measuring_2019}. 

In our setting, we probe the Hall response by measuring the drift along the synthetic dimension induced by an external force $F_x$ applied along $x$. This force is generated by introducing a time-dependent frequency detuning $\delta(t)$ between the Raman and lattice laser beams, producing an inertial force $F_x=\hbar\dot{\delta}/(Mk)$ along $x$~\cite{ben_dahan_bloch_1996}.

To uniformly sample the $2J+1$ subbands that constitute the ground band, we initialize the atoms in an incoherent mixture of all spin projections $m$ at quasi-momentum $q_x = 0$ (see Appendix~\ref{appendix:incoherent_mixture}). The Raman and lattice couplings are then ramped up over a duration of $\SI{300}{\micro\second}$ while keeping the tuning parameter $\lambda$ fixed. This timescale ensures that the loading process is adiabatic with respect to higher Bloch bands. Although the different subbands forming the ground band are not individually followed adiabatically, starting with an equally weighted initial distribution guarantees that all $2J+1$ subbands remain equally populated during the loading process.

We then apply the force $F_x$ for a duration $T=\SI{300}{\micro\second}$ using an S-shape detuning profile to ensure smooth acceleration and minimize diabatic excitations to higher bands. This detuning ramp corresponds  to one complete Bloch oscillation with quasi-momentum change $\Delta q_x=\pm2\hbar k$. After this evolution, we measure the spin-projection distribution $P(m)$ by imaging the atomic gas after a Stern--Gerlach separation along the vertical direction $z$.

\begin{figure}[t]
  \begin{center}
  \includegraphics[scale=0.92,trim=0 0 0 0]{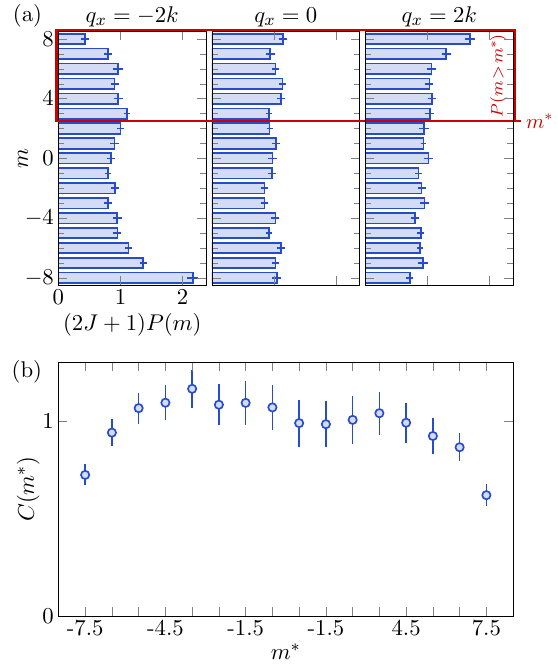}
  \end{center}
  \phantomsubfloat{\label{fig:Hall_measurement:Pim}}
  \phantomsubfloat{\label{fig:Hall_measurement:Cm}}
  \vspace{-3\baselineskip}
  \caption{
  \captionlabel{fig:Hall_measurement:Pim} Hall response measurement for $\lambda=-0.6$ and $U=4.8(2)\,\Erec$. The spin distributions $P(m)$ before ($q_x=0$) and after ($q_x=\pm2k$) one Bloch oscillation reveal the transfer of population across the synthetic dimension.
  \captionlabel{fig:Hall_measurement:Cm}
  Local Chern marker $C(m^*)$ from the variation of cumulative probability $P(m>m^*)$ [Eq.~\eqref{eq:Cmstar}]. Bulk values are consistent with unit Chern number $\mathcal{C}=1$. Error bars denote standard error over $\approx 40$ measurements.
  }
  \label{fig:Hall_measurement}
\end{figure}

In the topological phase, a Bloch oscillation induces population transfer across the synthetic dimension (Fig.~\ref{fig:Hall_measurement:Pim}). To quantify the Hall current through link $m^*$ (half-integer position separating the two adjacent spin states $m=m^*\pm1/2$), we measure the change in cumulative population $P(m>m^*)$. This variation defines the local Hall response at the link $m^*$, and defines the local Chern marker \cite{bianco_mapping_2011,chalopin_probing_2020} as:
\begin{equation}
C(m^*)=(2J+1)\frac{\Delta P(m>m^*)}{\Delta q_x/2\hbar k}.
\label{eq:Cmstar}
\end{equation}
In practice we perform Bloch oscillations along opposite directions $\Delta q_x=\pm2\hbar k$ and average the results. We show in \fig{Hall_measurement:Cm} an example of Chern marker $C(m^*)$ measured for $\lambda=-0.6$, i.e., in the topological phase. The Chern marker is nearly uniform and yields $C(m)=1.04(7)$ in a wide range $|m|\leq6$, consistent with quantized Chern number $\mathcal{C}=1$ in the bulk of the system.

\begin{figure}[th!]
  \begin{center}
  \includegraphics[scale=0.95,trim=10 0 0 0]{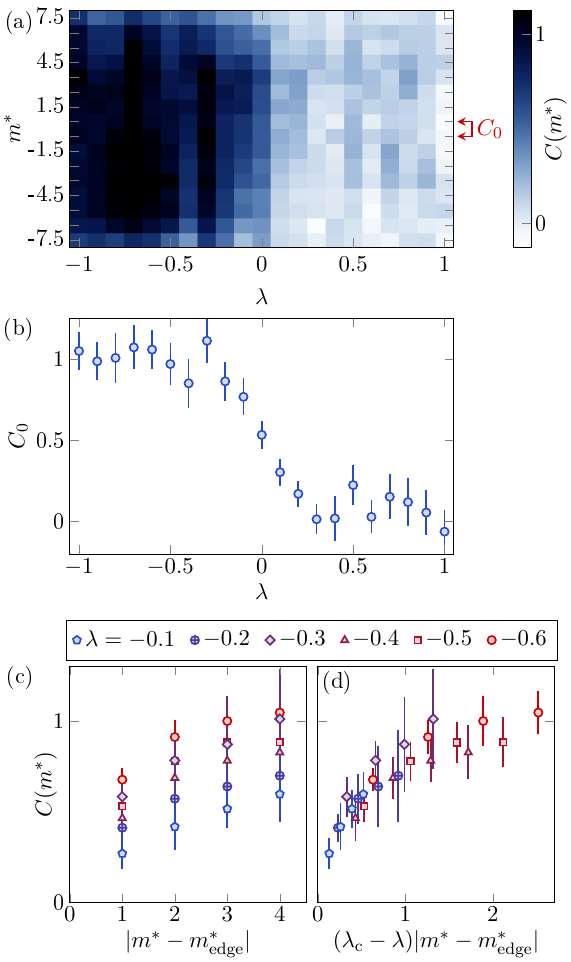}
  \end{center}
  \phantomsubfloat{\label{fig:hall_response:Cm}}
  \phantomsubfloat{\label{fig:hall_response:Cbulk}}
  \phantomsubfloat{\label{fig:hall_response:edge}}
  \phantomsubfloat{\label{fig:hall_response:edge_rescaled}}
  \vspace{-2\baselineskip}
  \caption{
\captionlabel{fig:hall_response:Cm} Local Chern marker $C(m^*)$ across the topological phase transition for $U=4.8(2)\,\Erec$. \captionlabel{fig:hall_response:Cbulk} Central Chern marker $C_0$ versus tuning parameter $\lambda$, showing robust quantization in gapped phases and rapid transition near $\lambda=0$. \captionlabel{fig:hall_response:edge} Chern marker decay near system edges (distance $|m^*_{\rm edge}|$) in the topological phase ($-0.6\leq\lambda<0$). \captionlabel{fig:hall_response:edge_rescaled} Same data with rescaled horizontal axis $(\lambda_{\mathrm{c}}-\lambda)|m^*-m^*_{\mathrm{edge}}|$, collapsing onto a universal curve with critical point $\lambda_{\mathrm{c}}=0.03(3)$. } \label{fig:hall_response} 
\end{figure}

The variation of the Chern marker across the topological phase transition is shown in \fig{hall_response:Cm}. We first focus on the bulk response, given by the central Chern marker $C_0$ defined as the average of $C(m)$ at central links $m^*=\pm1/2$ (see \fig{hall_response:Cbulk}).  Away from criticality, we observe robust quantization: $C_0=1.02(9)$ for $\lambda<-0.2$ and $C_0=0.06(10)$ for $\lambda>0.2$, matching expected Chern numbers $\mathcal{C}\in\{1,0\}$. Near the critical point ($|\lambda|\lesssim0.2$), the Chern marker rapidly transitions and becomes non-quantized, reflecting adiabaticity breakdown due to a closing of the bulk gap. The width of this transition region is limited by the finite duration of the Bloch oscillation as well as the finite size of the synthetic dimension. Nevertheless, the observed sharp variation of the Hall response near $\lambda=0$ provides a clear signature of the topological phase transition.

\subsection{Divergence of the edge correlation length\label{sec:edge_correlation_length}}

The topological phase transition is also reflected in the behavior of the edge modes. In the topological phase, the local Chern marker remains uniform in the bulk but decreases near the edges of the synthetic dimension (see \fig{hall_response:edge}). This reduction reflects the inability of the system to sustain a quantized Hall drift beyond its boundaries, giving rise to chiral edge modes with quasi-ballistic propagation along 
$x$~\cite{wen_chiral_1990}. The spatial region over which the Chern marker decays thus provides a direct measure of the edge-mode localization length.  

Approaching the topological phase transition, this localization length increases sharply, indicating progressive delocalization of the edge states into the bulk~\cite{bianco_mapping_2011,chen_correlation_2017,caio_topological_2019}. Focusing on the critical regime $-0.6 \le \lambda < 0$ in the topological phase, we show in \fig{hall_response:edge_rescaled} that the Chern-marker profiles collapse onto a single universal curve when plotted against the rescaled coordinate $(\lambda_{\mathrm{c}}-\lambda)|m^*-m^*_{\rm edge}|$. This scaling collapse demonstrates the divergence of a characteristic length scale $\xi \propto 1/|\lambda-\lambda_{\mathrm{c}}|$ at the transition~\cite{caio_topological_2019}. The optimal collapse is obtained for $\lambda_{\mathrm{c}}=0.03(3)$, consistent with the expected critical point $\lambda_{\mathrm{c}}=0$. Through the bulk-boundary correspondence of topological systems, the observed dilution of edge modes into the bulk constitutes a direct manifestation of the topological phase transition~\cite{qi_topological_2008}.

\subsection{Non-adiabatic dynamics and gap closing}

\begin{figure}[t]
  \begin{center}
  \includegraphics[scale=0.93,trim=5 0 0 0]{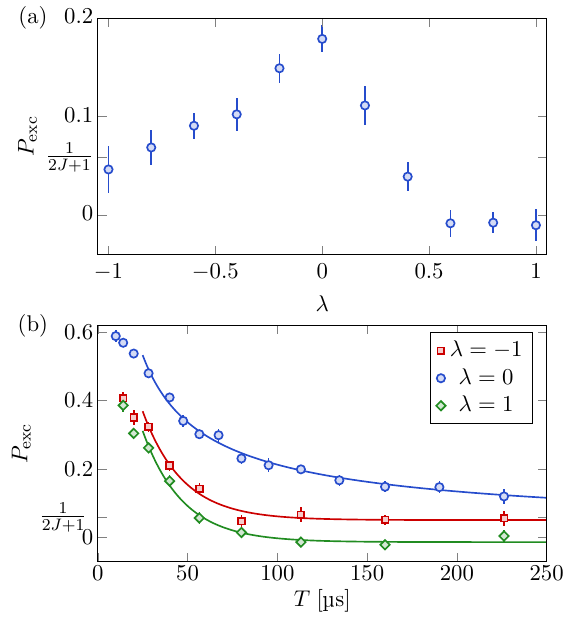}
  \end{center}
  \phantomsubfloat{\label{fig:adiabaticity:varylambda}}
  \phantomsubfloat{\label{fig:adiabaticity:varyT}}
  \vspace{-3\baselineskip}
  \caption{
  \captionlabel{fig:adiabaticity:varylambda} Fraction $P_{\text{exc}}$ of atoms excited to higher bands during a Bloch oscillation of duration $T=\SI{100}{\micro\second}$ as a function of the lattice tuning parameter $\lambda$.
  \captionlabel{fig:adiabaticity:varyT} Excited fraction $P_{\text{exc}}$ versus Bloch oscillation duration $T$ for $\lambda=-1,0,1$. Solid lines show fits in the range $\SI{25}{\micro\second}<T<\SI{250}{\micro\second}$: exponential with offset for $\lambda=\pm 1$, power law $\propto T^{-\alpha}$ for $\lambda=0$. The critical exponent $\alpha=0.65(5)$ is close to the expected value $\alpha=0.5$ for a two-dimensional Dirac point.
  }
  \label{fig:adiabaticity}
\end{figure}

The sharp variation of the Hall response near $\lambda=0$ and the delocalization of edge modes into the bulk signal a topological phase transition between a Chern insulator and a trivial phase. This transition requires the energy gap separating ground and excited bands to close~\cite{qi_topological_2008,thouless_quantization_1983}.

To locate the gap closing, we analyze Bloch-oscillation adiabaticity across the transition. We measure the fraction $P_{\text{exc}}$ of atoms excited to higher bands using standard band mapping, with a shortened oscillation duration $T=\SI{100}{\micro\second}$ to enhance excitations. Figure~\ref{fig:adiabaticity:varylambda} shows a sharp maximum in $P_{\text{exc}}$ at $\lambda=0$, directly indicating gap closure at the critical point.

Varying the Bloch oscillation duration $T$ at fixed $\lambda$ (Fig.~\ref{fig:adiabaticity:varyT}) reveals distinct dynamics across the phase diagram. In the trivial phase ($\lambda=1$), $P_{\text{exc}}$ decays exponentially with $T$, consistent with Landau--Zener suppression in gapped systems~\cite{landau_theory_1932,zener_non-adiabatic_1932}. The topological phase ($\lambda=-1$) shows a similarly rapid decay but toward a finite floor $P_{\text{exc}}^\infty=0.05(1)$, arising from gapless chiral edge modes~\cite{halperin_quantized_1982}: one subband connects ground and excited bands, undergoing diabatic transitions while the bulk modes remain adiabatic, yielding $P_{\text{exc}}^\infty\simeq 1/(2J+1)\approx 0.06$~\cite{hatsugai_chern_1993}.

At criticality ($\lambda=0$), $P_{\text{exc}}$ is substantially larger and decays as a power law $P_{\text{exc}}\propto T^{-\alpha}$ with exponent $\alpha=0.65(5)$. This scaling is close to the expected value $\alpha=1/2$ for a two-dimensional Dirac point~\cite{lim_bloch-zener_2012}, ruling out a quadratic band touching ($\alpha=1/4$). The slight deviation from $\alpha=1/2$ could be due to residual variations in the tuning parameter $\lambda$ across the synthetic dimension.

\section{A single Dirac point}

We now focus on the critical point$\lambda=0$ and characterize the nature of the gap closing.
The power-law scaling of band excitations upon Bloch oscillation already provided evidence for a Dirac-type gap closing (see \fig{adiabaticity:varyT}). However, this measurement does not distinguish between a single or multiple Dirac points. To resolve this question, we perform a series of momentum-space measurements that directly probe the local band structure around the gap closing point.

\begin{figure}[t]
  \begin{center}
  \includegraphics[scale=0.89, trim=10 0 0 0]{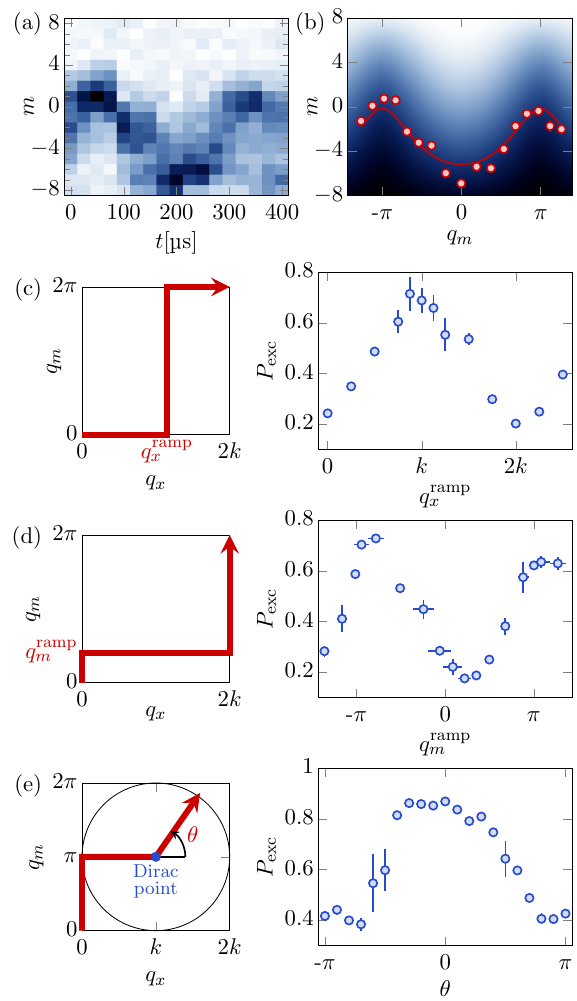}
  \end{center}
  \phantomsubfloat{\label{fig:Dirac_point:qm_BO}}
  \phantomsubfloat{\label{fig:Dirac_point:qm_BO_spin_model}}
  \phantomsubfloat{\label{fig:Dirac_point:Pexc_vs_qx}}
  \phantomsubfloat{\label{fig:Dirac_point:Pexc_vs_qm}}
  \phantomsubfloat{\label{fig:Dirac_point:winding_number}}
  \vspace{-3.8\baselineskip}
  \caption{
   \captionlabel{fig:Dirac_point:qm_BO} Synthetic-dimension Bloch oscillations at $\lambda=0$ induced by an effective Zeeman field with Larmor frequency $\nu_{\text{L}}=\SI{3}{\kilo\hertz}$, for a fixed quasi-momentum $q_x=0$.
    \captionlabel{fig:Dirac_point:qm_BO_spin_model} Comparison with quasi-classical spin dynamics showing that quasi-momentum $q_m$ traverses the full Brillouin zone over one evolution period. The color encodes the classical energy functional.
    \captionlabel{fig:Dirac_point:Pexc_vs_qx} Excited fraction $P_{\text{exc}}$ during $q_m$ Bloch oscillations versus initial $q_x$. The peak near $q_x=k$ locates the gap closing. 
    \captionlabel{fig:Dirac_point:Pexc_vs_qm} Excited fraction $P_{\text{exc}}$ during $q_x$ Bloch oscillations versus initial $q_m$. The peak near $q_m=\pi$ locates the gap closing.
    \captionlabel{fig:Dirac_point:winding_number} Excited fraction $P_{\text{exc}}$ after a two-segment trajectory in quasi-momentum space. Starting from the Dirac point along $q_x$, the system exits at angle $\theta$. The $\cos\theta$ variation of $P_{\text{exc}}(\theta)$ confirms unit winding around the Dirac point.
  }
  \label{fig:Dirac_point}
\end{figure}

\subsection{Manipulating quasi-momentum in 2D}

Resolving the band touching in momentum space requires access to both quasi-momenta $q_x$ and $q_m$. While $q_x$ is readily controlled via Bloch oscillations induced by a position-dependent force, accessing $q_m$--the azimuthal phase of the spin--is more challenging due to the finite extent of the synthetic dimension and open boundary conditions, which prevents $q_m$ from being a conserved quantity.

To overcome this limitation, we prepare quasi-momentum-localized wavepackets and manipulate their trajectories in two-dimensional quasi-momentum space $(q_x,q_m)$ on timescales short enough to suppress edge effects, so that $q_m$ is approximately conserved. We can then manipulate $q_m$  by applying an effective force along the synthetic dimension through an additional frequency detuning between the Raman laser beams, simulating a Zeeman field with Larmor frequency $\nu_\mathrm{L}$. As shown in \fig{Dirac_point:qm_BO}, the Zeeman field induces almost periodic oscillations of the spin distribution with a period equal to the Larmor period.

We validate this interpretation using an effective quasi-classical model (see Appendix~\ref{appendix:spin_functional}) in which spin operators are replaced by classical spin components. The excellent agreement between experiment and theory confirms that the observed oscillation corresponds to a genuine Bloch oscillation along the synthetic dimension, providing access to all quasi-momenta $q_m$ throughout the Brillouin zone.

\subsection{Locating the Dirac point in $(q_x,q_m)$ space}

Using this technique, we map the gap closing point in two-dimensional quasi-momentum space $(q_x,q_m)$ \cite{tarruell_creating_2012}. First, we measure the excited fraction $P_{\text{exc}}$ following a Bloch oscillation along $q_m$ of fixed duration $\SI{300}{\micro\second}$, for various initial $q_x$ values (Fig.~\ref{fig:Dirac_point:Pexc_vs_qx}). A pronounced peak near $q_x=k$ identifies the gap closing location along $q_x$. Similarly, performing Bloch oscillations along $q_x$ for various initial $q_m$ (using the same Bloch oscillation period) reveals a peak near $q_m=\pi$ mod $(2\pi)$ (Fig.~\ref{fig:Dirac_point:Pexc_vs_qm}), confirming the gap closing location along $q_m$. We conclude that the gap closes at a single point in the Brillouin zone, located at the $M$ point $(q_x,q_m)=(k,\pi)$, consistent with theoretical predictions.

\subsection{Topological charge of the Dirac point}

We probe the topological charge of the gap closing point to confirm its Dirac-type nature. Following the protocol of Ref.~\cite{brown_direct_2022}, we investigate the non-Abelian rotation of the effective spin-1/2 around the Dirac point by transporting momentum wavepackets across it. Specifically, the system is first accelerated along $q_x$ at $q_m=\pi$ to reach the gap closing point, then exits along a straight line at angle $\theta$ to the initial direction. In this experiment, both segments of the trajectory---toward and away from the Dirac point---use Raman and lattice detuning ramps that vary linearly in time, with a change of slope at the Dirac point. We measure the fraction of excited atoms $P_{\text{exc}}(\theta)$, which reflects the mismatch between the effective spin orientations for the incoming and outgoing momentum directions relative to the Dirac point. As shown in \fig{Dirac_point:winding_number}, we observe a clear $2\pi$ periodicity in $P_{\text{exc}}(\theta)$, consistent with unit topological charge of the Dirac point. This conclusively demonstrates the existence of a single Dirac point at criticality.

\section{Half-quantization and\\ parity anomaly}
When varying the tuning parameter $\lambda$ across the critical point, we observe a rapid transition in the Hall response from quantized value $C_0=1$ in the topological phase to $C_0=0$ in the trivial phase. At criticality ($\lambda=0$), we measure an intermediate value $C_0=0.53(9)$, consistent with half-quantization (see Fig.~\ref{fig:hall_response:Cbulk}). This half-quantized Hall response is a direct manifestation of the parity anomaly associated with the single Dirac point at criticality.

In order to assess the stability of this half-quantized Hall response, we subject the system to various perturbations that preserve the emergent parity symmetry of the Dirac Hamiltonian.

\begin{figure}[t]
  \begin{center}
  \includegraphics[scale=0.93,trim=10 0 0 0]{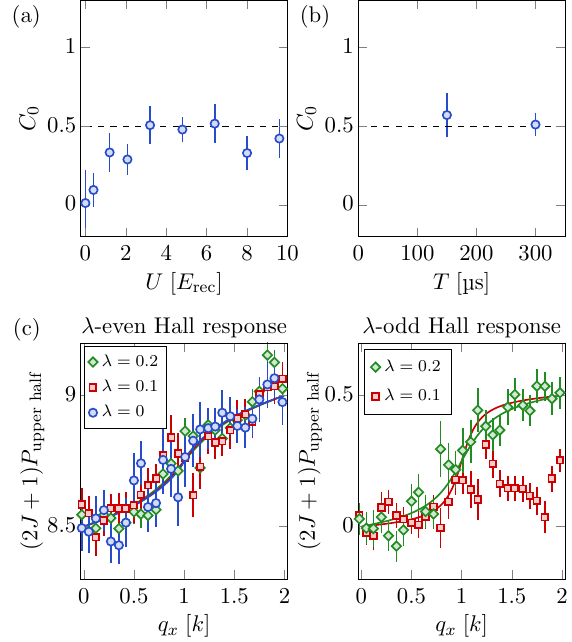}
  \end{center}
  \phantomsubfloat{\label{fig:Half_quantization:robustness_U}}
  \phantomsubfloat{\label{fig:Half_quantization:robustness_T}}
  \phantomsubfloat{\label{fig:Half_quantization:Berry_curvature}}
  \vspace{-3\baselineskip}
  \caption{
 \captionlabel{fig:Half_quantization:robustness_U}
Chern marker $C_0$ at criticality ($\lambda=0$) as a function of coupling strength $U$, demonstrating robustness to parity-preserving perturbations.
\captionlabel{fig:Half_quantization:robustness_T}
Chern marker $C_0$ versus Bloch oscillation duration $T$, showing stability against variations in adiabaticity.
\captionlabel{fig:Half_quantization:Berry_curvature}
Upper-half occupation probability $P_{\text{upper half}}$ decomposed into even and odd components under $\lambda \rightarrow -\lambda$, measured near criticality. Solid lines show calculations from the infinite coupled-wire model assuming an adiabatic Hall response. The measured even component is independent of $\lambda$ and contributes $C_0^{\text{even}} = 0.51(5)$ to the Chern marker, yielding half-quantization. The odd component sharpens near the Dirac point and restores integer quantization away from criticality ($|\lambda|=0.2$). Closer to criticality ($|\lambda|=0.1$), the measured odd contribution deviates from the theoretical prediction, indicating suppression of the Dirac-point contribution.
  }
  \label{fig:Half_quantization}
\end{figure}

We first vary the coupling strength $U$ at fixed criticality and Bloch oscillation period $T=\SI{300}{\micro\second}$ (see Fig.~\ref{fig:Half_quantization:robustness_U}).  In the range $3\,E_{\text{rec}}<U<10\,E_{\text{rec}}$, our measurements yield $C_0=0.45(5)$, confirming that half-quantization is robust to coupling strength variations. For small couplings $U\leq2.5\,E_{\text{rec}}$, the Chern marker $C_0$ deviates from half-quantization, which we attribute to the reduced bandgap and increased non-adiabatic excitations.

We also vary the duration of the Bloch oscillation between two values $T=\SI{150}{\micro\second}$ and $\SI{300}{\micro\second}$. While the band excitation probability varies by about 60\,\% (see Fig.~\ref{fig:adiabaticity:varyT}), the Hall response remains consistent with half-quantization for both durations  (see Fig.~\ref{fig:Half_quantization:robustness_T}). 


Since non-adiabatic transitions dominantly occur near the Dirac point, the robustness of half-quantization to the Bloch oscillation duration suggests it originates from quasi-momentum states away from the Dirac point. To test this, we measure the quasi-momentum-resolved Hall response near criticality.

We recall that the global bulk Hall response is quantified by the Chern marker $C_0$, defined as the increase of the occupation probability $P_{\text{upper half}}$ over one Bloch oscillation. Here $P_{\text{upper half}}\equiv 0.5\,P(m=0)+P(m>0)$. In this section, we examine the quasi-momentum dependence of $P_{\text{upper half}}$ across the Brillouin zone. This quantity is directly related to the Berry curvature $F(q_x,q_m)$ integrated over $q_m$ according to
\[
\langle F(q_x,q_m)\rangle_{q_m} = \frac{\partial P_{\text{upper half}}}{\partial q_x}.
\]

Following the theoretical analysis of the Berry curvature in the infinite coupled-wire model (see Sec.~\ref{sec:parity_anomaly}), we decompose $P_{\text{upper half}}$ into components that are even and odd under $\lambda \to -\lambda$. The even component is obtained as the average of the measurements at $\lambda$ and $-\lambda$, while the odd component corresponds to half their difference. Figure~\ref{fig:Half_quantization:Berry_curvature} displays the measured even and odd components of $P_{\text{upper half}}$ as a function of $q_x$ for several values of $\lambda$ near criticality, together with theoretical predictions from the infinite coupled-wire model assuming an adiabatic Hall response.

As shown in Fig.~\ref{fig:Half_quantization:Berry_curvature}, the even component remains nearly independent of $\lambda$ throughout the critical region and agrees well with the adiabatic prediction. Integrating this contribution yields $C_0^{\text{even}}=0.51(5)$, confirming half-quantization at the critical point $\lambda=0$, where the odd component vanishes by symmetry.

Further away from criticality ($|\lambda|=0.2$), the odd component also follows the adiabatic prediction, providing an additional half-quantized contribution $C_0^{\text{odd}}\approx\pm0.5$ to the global Hall response. The variation of $P_{\text{upper half}}$ with $q_x$ is most pronounced near $q_x\simeq k$, consistent with a concentration of the Berry curvature around the Dirac point. Repeating the measurement closer to criticality ($|\lambda|=0.1$), however, reveals a clear suppression of the Hall response relative to the adiabatic prediction, which we attribute to non-adiabatic transitions occurring in the vicinity of the Dirac point.

These observations highlight the gradual suppression of the Dirac-point contribution to the Hall response as the system approaches criticality, while the half-quantized contribution from states away from the Dirac point remains robust. Our measurements therefore provide a direct experimental signature of the parity anomaly through the emergence of a half-quantized Hall response at the critical point.

\section{Conclusion and outlook}

We have experimentally realized the parity anomaly in a genuinely two-dimensional quantum system by observing a half-quantized Hall response at the critical point of a topological phase transition. Using ultracold dysprosium atoms in a synthetic lattice, we engineered tunable Chern bands and identified a single Dirac point where the bulk gap closes. At this critical point, despite strong non-adiabatic excitations, the system exhibits a robust half-quantized Hall conductance—a direct manifestation of the fundamental incompatibility between gauge invariance and global parity symmetry at the quantum level. We established that this anomalous response originates from the global structure of the Brillouin zone and is protected by an emergent parity symmetry. Our results demonstrate a controlled realization of quantum anomalies in a purely two-dimensional setting, without relying on higher-dimensional anomaly inflow.

Beyond confirming the parity anomaly, our work reveals several striking features at the critical point:  power-law excitation scaling characteristic of Dirac physics, universal collapse of edge-mode profiles signaling divergent localization lengths. We further present in Appendix~\ref{appendix:coherence} a study of spatial coherence across the phase transition, showing extended coherence in the critical regime. These observations establish the synthetic lattice platform as a powerful testbed for exploring critical phenomena in topological systems.

Several directions merit future investigation. A central open question concerns the boundary response at parity-anomalous criticality. In contrast to fully gapped topological phases with well-defined edge states, a critical point with a half-quantized Hall response may exhibit exotic boundary behavior—including either chiral edge modes emanating from the Dirac point \cite{beenakker_chiral_2024} or algebraically decaying edge currents \cite{zou_half-quantized_2022}—fundamentally different from gapped phases. Recent observations of half-quantized chiral edge responses in three-dimensional magnetic topological insulators \cite{zhuo_half-quantized_2026} provide a first step towards the understanding of such edge physics. Our two-dimensional platform offers distinct advantages: direct spatial resolution of edge phenomena, local addressability via spin projections, and tunability of the Dirac point location in momentum space. These capabilities enable detailed mapping of boundary physics that would be difficult to access in three-dimensional systems.

Finally, extending this platform beyond the non-interacting regime opens up rich physics. While interactions are currently challenging due to atom loss from dipolar relaxation at high magnetic fields, this limitation can be overcome through lower-field operation \cite{lepoutre_out--equilibrium_2019,lecomte_production_2025} or tighter transverse confinement \cite{pasquiou_control_2010,barral_suppressing_2024}. In the interacting regime, theory predicts that interactions modify universal critical exponents of correlation functions near a single Dirac point \cite{chen_correlation_2017} and may drive spontaneous gapping through dynamical mass generation \cite{gross_dynamical_1974,tabatabaei_chiral_2022,gao_interacting_2025}. Exploring the competition between parity-anomalous Hall response and interaction-driven instabilities represents a compelling frontier for future work.

\section*{Acknowledgements}
We acknowledge insightful discussions with Baptiste Bermond, Anushya Chandran, Nathan Goldman, and David M. Long. We are especially thankful to Jérôme Beugnon and Jean Dalibard for their valuable comments and careful reading of the manuscript. We thank Jean-Baptiste Bouhiron, Evgenii Gadylshin and Qi Liu for their experimental assistance during the early stages of this project. This work is supported by the European Union (grant TOPODY 756722 from the European Research Council), the French Agence Nationale de la Recherche (grant HighDy ANR-24-CE47-2670), and Institut Universitaire de France. N.M. acknowledges support from DIM Quantip of Région Île-de-France.

\section*{Data availability}
The data that support the findings of this study are available in a Zenodo repository (DOI:10.5281/zenodo.19187224).

\appendix

\section{Experimental setup\label{appendix:experimental_setup}}

Our experiments begin with the preparation of an ultracold gas of approximately $9(1)\times 10^3$ $^{162}$Dy atoms confined in a crossed optical dipole trap with trapping frequencies $(\omega_{x'},\omega_{y'},\omega_z)=2\pi\times(73(5),147(5),200(10))\,\si{\hertz}$, where $x'$ and $y'$ are oriented at $\pm\SI{45}{\degree}$ relative to the $x$ axis defined by the Raman and lattice beams.

A bias magnetic field $B=\SI{197(1)}{\milli\gauss}$ is applied along the $z$-axis to define the quantization axis and establish a Zeeman splitting $\omega_z/2\pi=\SI{342(1)}{\kilo\hertz}$ between adjacent spin states. The atomic cloud is cooled to a temperature $T=\SI{103(5)}{\nano\kelvin}$, intentionally set slightly above the Bose-Einstein condensation critical temperature $T_c\simeq\SI{87}{\nano\kelvin}$. This temperature choice limits the peak atom density and suppresses atomic losses from dipolar relaxation over the experimental timescale, while being significantly lower then the recoil temperature $E_{\text{rec}}/k_B\simeq\SI{150}{\nano\kelvin}$.

The Raman and lattice laser beams share an elliptic polarization $\boldsymbol\epsilon=\sin\theta \ebold_y + \I \cos\theta\,\ebold_z$ with $\theta=\SI{32.9}{\degree}$. This polarization is chosen to ensure that Raman couplings follow the algebra of the spin-raising operator $J_+$ and that the lattice depth exhibits spin-dependent modulation $U_{\text{L}}[1-m^2/(2J^2)]$ (see \fig{implementation:couplings}). The Raman and lattice couplings are calibrated using Kapitza--Dirac diffraction resolved in spin space, yielding a precision of about 4\% on the tuning parameter $\lambda$. 

The vertical laser beam producing a tensor light shift $Q J_z^2$ is calibrated by measuring the one-axis twisting dynamics of an initially polarized spin state  \cite{kitagawa_squeezed_1993}. Its amplitude is then adjusted to the value that theoretically minimizes the width of the ground band for the given parameters $U$ and $\lambda$.

The atoms are adiabatically loaded into the ground band of the synthetic two-dimensional lattice by ramping the Raman and lattice laser intensities over \SI{300}{\micro\second}. At the end of the experimental sequence, an inverse ramp can be applied to return the system to the uncoupled regime for band-mapping measurements.

\section{Incoherent spin mixture\label{appendix:incoherent_mixture}}

To ensure uniform sampling of all $2J+1=17$ subbands forming the ground band, we prepare an incoherent mixture of all spin projections $m$ at quasi-momentum $q_x=0$ before loading into the synthetic lattice. This is accomplished using a sequence of 10 light pulses that induce nonlinear, chaotic spin dynamics, interlaced with radiofrequency $\pi/2$ pulses of randomized azimuthal angles. This procedure yields an approximately uniform spin distribution across all projections $m\in[-J,J]$.

A residual coherence between spin states results in a shot-to-shot variation $\delta P(m)=0.29\langle P(m)\rangle$. To mitigate this limitation on the precision of Chern-marker determination—which relies on measured values of $P(m)$—we repeat each measurement approximately 40 times. Under these conditions, the expected statistical error on $C(m)$ is approximately 0.13, consistent with our reported error bars.

\section{Effective spin functional\label{appendix:spin_functional}} 

To validate the interpretation of the observed oscillations as Bloch oscillations along the synthetic dimension (see \fig{Dirac_point:qm_BO}), we develop an effective quasi-classical model in which spin operators are replaced by classical spin components. 

The quantum spin operators $J_x$, $J_y$, and $J_z$ are replaced by classical variables $\sqrt{J^2-m^2}\cos\phi$, $\sqrt{J^2-m^2}\sin\phi$, and $m$, where $m$ is the spin projection along $z$ and $\phi$ is the azimuthal phase. These two quantities act as canonically conjugated variables with Poisson bracket $\{m,\phi\}=1$. Substituting these into the quantum Hamiltonian yields the effective classical Hamiltonian
\begin{align*}  
H_{\text{eff}}=\frac{p_x^2}{2M}&-U\bigg[(1-\lambda)\sqrt{1-m^2/J^2}\cos(2kx+\phi)\\
&+(1+\lambda)(1-m^2/2J^2)\cos(2kx)\bigg]+Qm^2.
\end{align*}
The Raman and lattice potentials combine into a single effective lattice potential
\[
V_{\text{eff}}(m,\phi)=U_{\text{eff}}(m,\phi)\cos[2kx+\varphi(m,\phi)],
\]
where the amplitude and phase depend explicitly on the spin variables. To extract the spin dynamics, we exploit the separation of timescales between fast kinetic motion along $x$ and slow spin evolution. For each spin configuration $(m,\phi)$, the system adiabatically follows the lowest Bloch state of $V_{\text{eff}}(m,\phi)$, acquiring a ground-band energy $E_0(m,\phi)$ evaluated at quasi-momentum $q_x=0$. This adiabatic elimination yields an effective spin functional
\[ 
\mathcal{E}(m,\phi)=E_0(m,\phi)+Q m^2. 
\] 
The classical equations of motion $\dot{m}=-\partial_\phi \mathcal{E}$ and $\dot{\phi}=\partial_m \mathcal{E}$ then govern spin dynamics. Numerical integration produces quasi-momentum trajectories in excellent agreement with experimental observations (see \fig{Dirac_point:qm_BO_spin_model}), validating the quasi-classical picture and confirming that the observed spin oscillations indeed represent Bloch oscillations along the synthetic dimension.

\section{Extended coherence at criticality\label{appendix:coherence}}

The presence of a Dirac point at criticality is expected to substantially alter the spatial coherence properties of the system. We probe these properties using spin- and velocity-resolved measurements of the atomic distribution $\rho(v_x,m)$ averaged over the ground band. Spatial coherence is inferred from the Fourier transform along $x$, yielding the first-order correlation function $g_1(\Delta x,m)$.

\begin{figure}[t]
  \begin{center}
  \includegraphics[scale=0.93,trim=10 0 0 0]{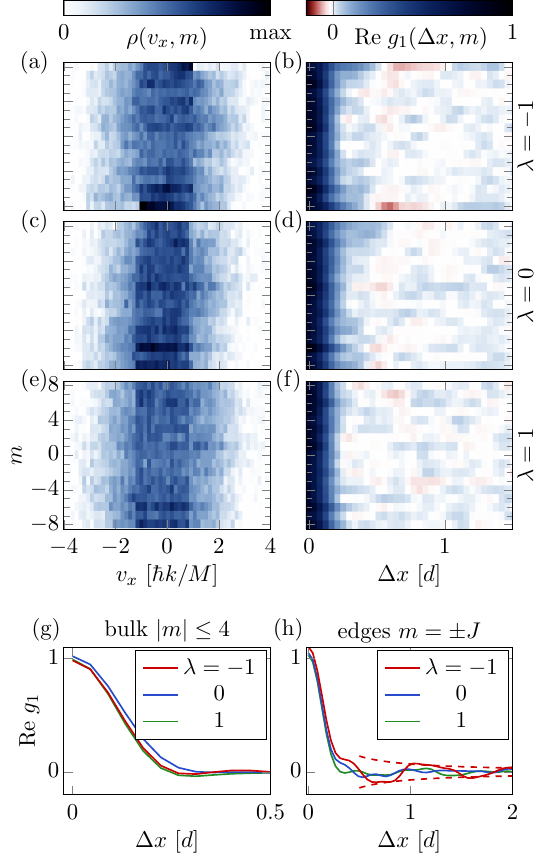}
  \end{center}
  \phantomsubfloat{\label{fig:coherence:velocity_lambdam1}}
  \phantomsubfloat{\label{fig:coherence:g1_lambdam1}}
  \phantomsubfloat{\label{fig:coherence:velocity_lambda0}}
  \phantomsubfloat{\label{fig:coherence:g1_lambda0}}
  \phantomsubfloat{\label{fig:coherence:velocity_lambda1}}
  \phantomsubfloat{\label{fig:coherence:g1_lambda1}}
  \phantomsubfloat{\label{fig:coherence:g1_bulk}}
  \phantomsubfloat{\label{fig:coherence:g1_edge}}
  \vspace{-3\baselineskip}
  \caption{
  Spin-resolved velocity distributions $\rho(v_x,m)$ and first-order correlation functions $g_1(m,\Delta x)$ for three representative phases.
  \captionlabel{fig:coherence:velocity_lambdam1} Velocity distribution in the topological phase ($\lambda=-1$): smooth bulk with chiral edge modes. 
  \captionlabel{fig:coherence:g1_lambdam1} The bulk exhibits short-range coherence, whereas the edges show a long-range algebraic decay.
  \captionlabel{fig:coherence:velocity_lambda0} At criticality ($\lambda=0$), the velocity distribution sharpens with an incipient cusp at $\pm\hbar k/M$.
  \captionlabel{fig:coherence:g1_lambda0} Correlation function broadens significantly, consistent with Dirac physics.
  \captionlabel{fig:coherence:velocity_lambda1} In the trivial phase ($\lambda=1$), the distribution remains smooth.
  \captionlabel{fig:coherence:g1_lambda1} Short-range coherence is recovered throughout.
  \captionlabel{fig:coherence:g1_bulk} Correlation function $g_1(\Delta x)$ averaged over the bulk $|m|\leq4$. At criticality, the coherence envelope slightly broadens compared to gapped phases, consistent with extended correlations.
  \captionlabel{fig:coherence:g1_edge} Correlation function at the edges $m=\pm J$ In the topological phase, it exhibits an algebraic decay consistent with chiral Luttinger-liquid behavior. Dashed red lines are $\propto\pm1/\Delta x$ guides to the eye.
  }
  \label{fig:coherence}
\end{figure}

In the topological phase ($\lambda<0$), bulk and edge regions exhibit markedly different coherence (see \fig{coherence}a,b). The bulk velocity distribution is smooth, yielding short-range coherence. Conversely, the edges ($m=\pm J$) exhibit sharp discontinuities at $v_x=\pm\vrec$, where $\vrec=\hbar k/M$ is the recoil velocity. This discontinuity leads to algebraic decay of the correlation function $g_1(\Delta x,m=\pm J)\sim e^{ik\Delta x}/|\Delta x|$ characteristic of chiral Luttinger-liquid behavior \cite{wen_chiral_1990} (see \fig{coherence}h). Conversely, the trivial phase ($\lambda>0$) exhibits smooth distributions and short-range coherence throughout  (see \fig{coherence}e,f). In the bulk of both phases ($|m|\leq4$), the velocity distribution exhibits very similar shapes with rms velocity width $\sigma_{v_x}=1.45(4)\,\vrec$ and $1.49(3)\,\vrec$ for $\lambda=-1$ and +1, respectively. This indicates that the bulk coherence properties are largely insensitive to the topological nature of the phase. 

At criticality ($\lambda=0$), the gap closing fundamentally alters the coherence structure. One expects the bulk velocity distribution to sharpen with an incipient logarithmic cusp at $\pm\vrec$, signaling extended coherence compared to gapped phases. The correlation function is expected to decay algebraically as $g_1(\Delta x) \sim 1/(\Delta x)^2$, characteristic of a two-dimensional Dirac fermion \cite{peskin_introduction_2018,sachdev_quantum_2011}. While experimental noise limits the resolution of asymptotic behavior, we observed a sharpening of the velocity distribution in the bulk (rms velocity width $\sigma_{v_x}=1.35(5)\,\vrec$) and a corresponding broadening of the coherence envelope, consistent with  extended correlations at criticality (see \fig{coherence}c,d,g).

\begin{thebibliography}{69}%
\makeatletter
\providecommand \@ifxundefined [1]{%
 \@ifx{#1\undefined}
}%
\providecommand \@ifnum [1]{%
 \ifnum #1\expandafter \@firstoftwo
 \else \expandafter \@secondoftwo
 \fi
}%
\providecommand \@ifx [1]{%
 \ifx #1\expandafter \@firstoftwo
 \else \expandafter \@secondoftwo
 \fi
}%
\providecommand \natexlab [1]{#1}%
\providecommand \enquote  [1]{``#1''}%
\providecommand \bibnamefont  [1]{#1}%
\providecommand \bibfnamefont [1]{#1}%
\providecommand \citenamefont [1]{#1}%
\providecommand \href@noop [0]{\@secondoftwo}%
\providecommand \href [0]{\begingroup \@sanitize@url \@href}%
\providecommand \@href[1]{\@@startlink{#1}\@@href}%
\providecommand \@@href[1]{\endgroup#1\@@endlink}%
\providecommand \@sanitize@url [0]{\catcode `\\12\catcode `\$12\catcode
  `\&12\catcode `\#12\catcode `\^12\catcode `\_12\catcode `\%12\relax}%
\providecommand \@@startlink[1]{}%
\providecommand \@@endlink[0]{}%
\providecommand \url  [0]{\begingroup\@sanitize@url \@url }%
\providecommand \@url [1]{\endgroup\@href {#1}{\urlprefix }}%
\providecommand \urlprefix  [0]{URL }%
\providecommand \Eprint [0]{\href }%
\providecommand \doibase [0]{https://doi.org/}%
\providecommand \selectlanguage [0]{\@gobble}%
\providecommand \bibinfo  [0]{\@secondoftwo}%
\providecommand \bibfield  [0]{\@secondoftwo}%
\providecommand \translation [1]{[#1]}%
\providecommand \BibitemOpen [0]{}%
\providecommand \bibitemStop [0]{}%
\providecommand \bibitemNoStop [0]{.\EOS\space}%
\providecommand \EOS [0]{\spacefactor3000\relax}%
\providecommand \BibitemShut  [1]{\csname bibitem#1\endcsname}%
\let\auto@bib@innerbib\@empty
\bibitem [{\citenamefont {Bertlmann}(2000)}]{bertlmann_anomalies_2000}%
  \BibitemOpen
  \bibfield  {author} {\bibinfo {author} {\bibfnamefont {R.~A.}\ \bibnamefont
  {Bertlmann}},\ }\href@noop {} {\emph {\bibinfo {title} {Anomalies in
  {{Quantum Field Theory}}}}},\ International {{Series}} of {{Monographs}} on
  {{Physics}}\ (\bibinfo  {publisher} {Oxford University Press},\ \bibinfo
  {address} {Oxford, New York},\ \bibinfo {year} {2000})\BibitemShut {NoStop}%
\bibitem [{\citenamefont {Fradkin}(2013)}]{fradkin_field_2013}%
  \BibitemOpen
  \bibfield  {author} {\bibinfo {author} {\bibfnamefont {E.}~\bibnamefont
  {Fradkin}},\ }\href {https://doi.org/10.1017/CBO9781139015509} {\emph
  {\bibinfo {title} {Field {{Theories}} of {{Condensed Matter Physics}}}}},\
  \bibinfo {edition} {2nd}\ ed.\ (\bibinfo  {publisher} {Cambridge University
  Press},\ \bibinfo {address} {Cambridge},\ \bibinfo {year} {2013})\BibitemShut
  {NoStop}%
\bibitem [{\citenamefont {Adler}(1969)}]{adler_axial-vector_1969}%
  \BibitemOpen
  \bibfield  {author} {\bibinfo {author} {\bibfnamefont {S.~L.}\ \bibnamefont
  {Adler}},\ }\href {https://doi.org/10.1103/PhysRev.177.2426} {\bibfield
  {journal} {\bibinfo  {journal} {Phys. Rev.}\ }\textbf {\bibinfo {volume}
  {177}},\ \bibinfo {pages} {2426} (\bibinfo {year} {1969})}\BibitemShut
  {NoStop}%
\bibitem [{\citenamefont {Bell}\ and\ \citenamefont
  {Jackiw}(1969)}]{bell_pcac_1969}%
  \BibitemOpen
  \bibfield  {author} {\bibinfo {author} {\bibfnamefont {J.~S.}\ \bibnamefont
  {Bell}}\ and\ \bibinfo {author} {\bibfnamefont {R.}~\bibnamefont {Jackiw}},\
  }\href {https://doi.org/10.1007/BF02823296} {\bibfield  {journal} {\bibinfo
  {journal} {Nuovo Cimento A (1965-1970)}\ }\textbf {\bibinfo {volume} {60}},\
  \bibinfo {pages} {47} (\bibinfo {year} {1969})}\BibitemShut {NoStop}%
\bibitem [{\citenamefont {Nielsen}\ and\ \citenamefont
  {Ninomiya}(1983)}]{nielsen_adler-bell-jackiw_1983}%
  \BibitemOpen
  \bibfield  {author} {\bibinfo {author} {\bibfnamefont {H.~B.}\ \bibnamefont
  {Nielsen}}\ and\ \bibinfo {author} {\bibfnamefont {M.}~\bibnamefont
  {Ninomiya}},\ }\href {https://doi.org/10.1016/0370-2693(83)91529-0}
  {\bibfield  {journal} {\bibinfo  {journal} {Physics Letters B}\ }\textbf
  {\bibinfo {volume} {130}},\ \bibinfo {pages} {389} (\bibinfo {year}
  {1983})}\BibitemShut {NoStop}%
\bibitem [{\citenamefont {Xiong}\ \emph {et~al.}(2015)\citenamefont {Xiong},
  \citenamefont {Kushwaha}, \citenamefont {Liang}, \citenamefont {Krizan},
  \citenamefont {Hirschberger}, \citenamefont {Wang}, \citenamefont {Cava},\
  and\ \citenamefont {Ong}}]{xiong_evidence_2015}%
  \BibitemOpen
  \bibfield  {author} {\bibinfo {author} {\bibfnamefont {J.}~\bibnamefont
  {Xiong}}, \bibinfo {author} {\bibfnamefont {S.~K.}\ \bibnamefont {Kushwaha}},
  \bibinfo {author} {\bibfnamefont {T.}~\bibnamefont {Liang}}, \bibinfo
  {author} {\bibfnamefont {J.~W.}\ \bibnamefont {Krizan}}, \bibinfo {author}
  {\bibfnamefont {M.}~\bibnamefont {Hirschberger}}, \bibinfo {author}
  {\bibfnamefont {W.}~\bibnamefont {Wang}}, \bibinfo {author} {\bibfnamefont
  {R.~J.}\ \bibnamefont {Cava}},\ and\ \bibinfo {author} {\bibfnamefont
  {N.~P.}\ \bibnamefont {Ong}},\ }\href
  {https://doi.org/10.1126/science.aac6089} {\bibfield  {journal} {\bibinfo
  {journal} {Science}\ }\textbf {\bibinfo {volume} {350}},\ \bibinfo {pages}
  {413} (\bibinfo {year} {2015})}\BibitemShut {NoStop}%
\bibitem [{\citenamefont {Huang}\ \emph {et~al.}(2015)\citenamefont {Huang},
  \citenamefont {Zhao}, \citenamefont {Long}, \citenamefont {Wang},
  \citenamefont {Chen}, \citenamefont {Yang}, \citenamefont {Liang},
  \citenamefont {Xue}, \citenamefont {Weng}, \citenamefont {Fang},
  \citenamefont {Dai},\ and\ \citenamefont {Chen}}]{huang_observation_2015}%
  \BibitemOpen
  \bibfield  {author} {\bibinfo {author} {\bibfnamefont {X.}~\bibnamefont
  {Huang}}, \bibinfo {author} {\bibfnamefont {L.}~\bibnamefont {Zhao}},
  \bibinfo {author} {\bibfnamefont {Y.}~\bibnamefont {Long}}, \bibinfo {author}
  {\bibfnamefont {P.}~\bibnamefont {Wang}}, \bibinfo {author} {\bibfnamefont
  {D.}~\bibnamefont {Chen}}, \bibinfo {author} {\bibfnamefont {Z.}~\bibnamefont
  {Yang}}, \bibinfo {author} {\bibfnamefont {H.}~\bibnamefont {Liang}},
  \bibinfo {author} {\bibfnamefont {M.}~\bibnamefont {Xue}}, \bibinfo {author}
  {\bibfnamefont {H.}~\bibnamefont {Weng}}, \bibinfo {author} {\bibfnamefont
  {Z.}~\bibnamefont {Fang}}, \bibinfo {author} {\bibfnamefont {X.}~\bibnamefont
  {Dai}},\ and\ \bibinfo {author} {\bibfnamefont {G.}~\bibnamefont {Chen}},\
  }\href {https://doi.org/10.1103/PhysRevX.5.031023} {\bibfield  {journal}
  {\bibinfo  {journal} {Phys. Rev. X}\ }\textbf {\bibinfo {volume} {5}},\
  \bibinfo {pages} {031023} (\bibinfo {year} {2015})}\BibitemShut {NoStop}%
\bibitem [{\citenamefont {Zhang}\ \emph {et~al.}(2016)\citenamefont {Zhang},
  \citenamefont {Xu}, \citenamefont {Belopolski}, \citenamefont {Yuan},
  \citenamefont {Lin}, \citenamefont {Tong}, \citenamefont {Bian},
  \citenamefont {Alidoust}, \citenamefont {Lee}, \citenamefont {Huang},
  \citenamefont {Chang}, \citenamefont {Chang}, \citenamefont {Hsu},
  \citenamefont {Jeng}, \citenamefont {Neupane}, \citenamefont {Sanchez},
  \citenamefont {Zheng}, \citenamefont {Wang}, \citenamefont {Lin},
  \citenamefont {Zhang}, \citenamefont {Lu}, \citenamefont {Shen},
  \citenamefont {Neupert}, \citenamefont {Zahid~Hasan},\ and\ \citenamefont
  {Jia}}]{zhang_signatures_2016}%
  \BibitemOpen
  \bibfield  {author} {\bibinfo {author} {\bibfnamefont {C.-L.}\ \bibnamefont
  {Zhang}}, \bibinfo {author} {\bibfnamefont {S.-Y.}\ \bibnamefont {Xu}},
  \bibinfo {author} {\bibfnamefont {I.}~\bibnamefont {Belopolski}}, \bibinfo
  {author} {\bibfnamefont {Z.}~\bibnamefont {Yuan}}, \bibinfo {author}
  {\bibfnamefont {Z.}~\bibnamefont {Lin}}, \bibinfo {author} {\bibfnamefont
  {B.}~\bibnamefont {Tong}}, \bibinfo {author} {\bibfnamefont {G.}~\bibnamefont
  {Bian}}, \bibinfo {author} {\bibfnamefont {N.}~\bibnamefont {Alidoust}},
  \bibinfo {author} {\bibfnamefont {C.-C.}\ \bibnamefont {Lee}}, \bibinfo
  {author} {\bibfnamefont {S.-M.}\ \bibnamefont {Huang}}, \bibinfo {author}
  {\bibfnamefont {T.-R.}\ \bibnamefont {Chang}}, \bibinfo {author}
  {\bibfnamefont {G.}~\bibnamefont {Chang}}, \bibinfo {author} {\bibfnamefont
  {C.-H.}\ \bibnamefont {Hsu}}, \bibinfo {author} {\bibfnamefont {H.-T.}\
  \bibnamefont {Jeng}}, \bibinfo {author} {\bibfnamefont {M.}~\bibnamefont
  {Neupane}}, \bibinfo {author} {\bibfnamefont {D.~S.}\ \bibnamefont
  {Sanchez}}, \bibinfo {author} {\bibfnamefont {H.}~\bibnamefont {Zheng}},
  \bibinfo {author} {\bibfnamefont {J.}~\bibnamefont {Wang}}, \bibinfo {author}
  {\bibfnamefont {H.}~\bibnamefont {Lin}}, \bibinfo {author} {\bibfnamefont
  {C.}~\bibnamefont {Zhang}}, \bibinfo {author} {\bibfnamefont {H.-Z.}\
  \bibnamefont {Lu}}, \bibinfo {author} {\bibfnamefont {S.-Q.}\ \bibnamefont
  {Shen}}, \bibinfo {author} {\bibfnamefont {T.}~\bibnamefont {Neupert}},
  \bibinfo {author} {\bibfnamefont {M.}~\bibnamefont {Zahid~Hasan}},\ and\
  \bibinfo {author} {\bibfnamefont {S.}~\bibnamefont {Jia}},\ }\href
  {https://doi.org/10.1038/ncomms10735} {\bibfield  {journal} {\bibinfo
  {journal} {Nat Commun}\ }\textbf {\bibinfo {volume} {7}},\ \bibinfo {pages}
  {10735} (\bibinfo {year} {2016})}\BibitemShut {NoStop}%
\bibitem [{\citenamefont {Ong}\ and\ \citenamefont
  {Liang}(2021)}]{ong_experimental_2021}%
  \BibitemOpen
  \bibfield  {author} {\bibinfo {author} {\bibfnamefont {N.~P.}\ \bibnamefont
  {Ong}}\ and\ \bibinfo {author} {\bibfnamefont {S.}~\bibnamefont {Liang}},\
  }\href {https://doi.org/10.1038/s42254-021-00310-9} {\bibfield  {journal}
  {\bibinfo  {journal} {Nat Rev Phys}\ }\textbf {\bibinfo {volume} {3}},\
  \bibinfo {pages} {394} (\bibinfo {year} {2021})}\BibitemShut {NoStop}%
\bibitem [{\citenamefont {Callan}\ and\ \citenamefont
  {Harvey}(1985)}]{callan_anomalies_1985}%
  \BibitemOpen
  \bibfield  {author} {\bibinfo {author} {\bibfnamefont {C.~G.}\ \bibnamefont
  {Callan}}\ and\ \bibinfo {author} {\bibfnamefont {J.~A.}\ \bibnamefont
  {Harvey}},\ }\href {https://doi.org/10.1016/0550-3213(85)90489-4} {\bibfield
  {journal} {\bibinfo  {journal} {Nuclear Physics B}\ }\textbf {\bibinfo
  {volume} {250}},\ \bibinfo {pages} {427} (\bibinfo {year}
  {1985})}\BibitemShut {NoStop}%
\bibitem [{\citenamefont {Qi}\ \emph {et~al.}(2008)\citenamefont {Qi},
  \citenamefont {Hughes},\ and\ \citenamefont {Zhang}}]{qi_topological_2008}%
  \BibitemOpen
  \bibfield  {author} {\bibinfo {author} {\bibfnamefont {X.-L.}\ \bibnamefont
  {Qi}}, \bibinfo {author} {\bibfnamefont {T.~L.}\ \bibnamefont {Hughes}},\
  and\ \bibinfo {author} {\bibfnamefont {S.-C.}\ \bibnamefont {Zhang}},\ }\href
  {https://doi.org/10.1103/PhysRevB.78.195424} {\bibfield  {journal} {\bibinfo
  {journal} {Phys. Rev. B}\ }\textbf {\bibinfo {volume} {78}},\ \bibinfo
  {pages} {195424} (\bibinfo {year} {2008})}\BibitemShut {NoStop}%
\bibitem [{\citenamefont {Niemi}\ and\ \citenamefont
  {Semenoff}(1983)}]{niemi_axial-anomaly-induced_1983}%
  \BibitemOpen
  \bibfield  {author} {\bibinfo {author} {\bibfnamefont {A.~J.}\ \bibnamefont
  {Niemi}}\ and\ \bibinfo {author} {\bibfnamefont {G.~W.}\ \bibnamefont
  {Semenoff}},\ }\href {https://doi.org/10.1103/PhysRevLett.51.2077} {\bibfield
   {journal} {\bibinfo  {journal} {Phys. Rev. Lett.}\ }\textbf {\bibinfo
  {volume} {51}},\ \bibinfo {pages} {2077} (\bibinfo {year}
  {1983})}\BibitemShut {NoStop}%
\bibitem [{\citenamefont {Redlich}(1984)}]{redlich_parity_1984}%
  \BibitemOpen
  \bibfield  {author} {\bibinfo {author} {\bibfnamefont {A.~N.}\ \bibnamefont
  {Redlich}},\ }\href {https://doi.org/10.1103/PhysRevD.29.2366} {\bibfield
  {journal} {\bibinfo  {journal} {Phys. Rev. D}\ }\textbf {\bibinfo {volume}
  {29}},\ \bibinfo {pages} {2366} (\bibinfo {year} {1984})}\BibitemShut
  {NoStop}%
\bibitem [{\citenamefont {Haldane}(1988)}]{haldane_model_1988}%
  \BibitemOpen
  \bibfield  {author} {\bibinfo {author} {\bibfnamefont {F.~D.~M.}\
  \bibnamefont {Haldane}},\ }\href
  {https://doi.org/10.1103/PhysRevLett.61.2015} {\bibfield  {journal} {\bibinfo
   {journal} {Phys. Rev. Lett.}\ }\textbf {\bibinfo {volume} {61}},\ \bibinfo
  {pages} {2015} (\bibinfo {year} {1988})}\BibitemShut {NoStop}%
\bibitem [{\citenamefont {Nielsen}\ and\ \citenamefont
  {Ninomiya}(1981)}]{nielsen_no-go_1981}%
  \BibitemOpen
  \bibfield  {author} {\bibinfo {author} {\bibfnamefont {H.~B.}\ \bibnamefont
  {Nielsen}}\ and\ \bibinfo {author} {\bibfnamefont {M.}~\bibnamefont
  {Ninomiya}},\ }\href {https://doi.org/10.1016/0370-2693(81)91026-1}
  {\bibfield  {journal} {\bibinfo  {journal} {Phys. Lett. B}\ }\textbf
  {\bibinfo {volume} {105}},\ \bibinfo {pages} {219} (\bibinfo {year}
  {1981})}\BibitemShut {NoStop}%
\bibitem [{\citenamefont {Novoselov}\ \emph {et~al.}(2005)\citenamefont
  {Novoselov}, \citenamefont {Geim}, \citenamefont {Morozov}, \citenamefont
  {Jiang}, \citenamefont {Katsnelson}, \citenamefont {Grigorieva},
  \citenamefont {Dubonos},\ and\ \citenamefont
  {Firsov}}]{novoselov_two-dimensional_2005}%
  \BibitemOpen
  \bibfield  {author} {\bibinfo {author} {\bibfnamefont {K.~S.}\ \bibnamefont
  {Novoselov}}, \bibinfo {author} {\bibfnamefont {A.~K.}\ \bibnamefont {Geim}},
  \bibinfo {author} {\bibfnamefont {S.~V.}\ \bibnamefont {Morozov}}, \bibinfo
  {author} {\bibfnamefont {D.}~\bibnamefont {Jiang}}, \bibinfo {author}
  {\bibfnamefont {M.~I.}\ \bibnamefont {Katsnelson}}, \bibinfo {author}
  {\bibfnamefont {I.~V.}\ \bibnamefont {Grigorieva}}, \bibinfo {author}
  {\bibfnamefont {S.~V.}\ \bibnamefont {Dubonos}},\ and\ \bibinfo {author}
  {\bibfnamefont {A.~A.}\ \bibnamefont {Firsov}},\ }\href
  {https://doi.org/10.1038/nature04233} {\bibfield  {journal} {\bibinfo
  {journal} {Nature}\ }\textbf {\bibinfo {volume} {438}},\ \bibinfo {pages}
  {197} (\bibinfo {year} {2005})}\BibitemShut {NoStop}%
\bibitem [{\citenamefont {Yu}\ \emph {et~al.}(2010)\citenamefont {Yu},
  \citenamefont {Zhang}, \citenamefont {Zhang}, \citenamefont {Zhang},
  \citenamefont {Dai},\ and\ \citenamefont {Fang}}]{yu_quantized_2010}%
  \BibitemOpen
  \bibfield  {author} {\bibinfo {author} {\bibfnamefont {R.}~\bibnamefont
  {Yu}}, \bibinfo {author} {\bibfnamefont {W.}~\bibnamefont {Zhang}}, \bibinfo
  {author} {\bibfnamefont {H.-J.}\ \bibnamefont {Zhang}}, \bibinfo {author}
  {\bibfnamefont {S.-C.}\ \bibnamefont {Zhang}}, \bibinfo {author}
  {\bibfnamefont {X.}~\bibnamefont {Dai}},\ and\ \bibinfo {author}
  {\bibfnamefont {Z.}~\bibnamefont {Fang}},\ }\href
  {https://doi.org/10.1126/science.1187485} {\bibfield  {journal} {\bibinfo
  {journal} {Science}\ }\textbf {\bibinfo {volume} {329}},\ \bibinfo {pages}
  {61} (\bibinfo {year} {2010})}\BibitemShut {NoStop}%
\bibitem [{\citenamefont {Chang}\ \emph {et~al.}(2013)\citenamefont {Chang},
  \citenamefont {Zhang}, \citenamefont {Feng}, \citenamefont {Shen},
  \citenamefont {Zhang}, \citenamefont {Guo}, \citenamefont {Li}, \citenamefont
  {Ou}, \citenamefont {Wei}, \citenamefont {Wang}, \citenamefont {Ji},
  \citenamefont {Feng}, \citenamefont {Ji}, \citenamefont {Chen}, \citenamefont
  {Jia}, \citenamefont {Dai}, \citenamefont {Fang}, \citenamefont {Zhang},
  \citenamefont {He}, \citenamefont {Wang}, \citenamefont {Lu}, \citenamefont
  {Ma},\ and\ \citenamefont {Xue}}]{chang_experimental_2013}%
  \BibitemOpen
  \bibfield  {author} {\bibinfo {author} {\bibfnamefont {C.-Z.}\ \bibnamefont
  {Chang}}, \bibinfo {author} {\bibfnamefont {J.}~\bibnamefont {Zhang}},
  \bibinfo {author} {\bibfnamefont {X.}~\bibnamefont {Feng}}, \bibinfo {author}
  {\bibfnamefont {J.}~\bibnamefont {Shen}}, \bibinfo {author} {\bibfnamefont
  {Z.}~\bibnamefont {Zhang}}, \bibinfo {author} {\bibfnamefont
  {M.}~\bibnamefont {Guo}}, \bibinfo {author} {\bibfnamefont {K.}~\bibnamefont
  {Li}}, \bibinfo {author} {\bibfnamefont {Y.}~\bibnamefont {Ou}}, \bibinfo
  {author} {\bibfnamefont {P.}~\bibnamefont {Wei}}, \bibinfo {author}
  {\bibfnamefont {L.-L.}\ \bibnamefont {Wang}}, \bibinfo {author}
  {\bibfnamefont {Z.-Q.}\ \bibnamefont {Ji}}, \bibinfo {author} {\bibfnamefont
  {Y.}~\bibnamefont {Feng}}, \bibinfo {author} {\bibfnamefont {S.}~\bibnamefont
  {Ji}}, \bibinfo {author} {\bibfnamefont {X.}~\bibnamefont {Chen}}, \bibinfo
  {author} {\bibfnamefont {J.}~\bibnamefont {Jia}}, \bibinfo {author}
  {\bibfnamefont {X.}~\bibnamefont {Dai}}, \bibinfo {author} {\bibfnamefont
  {Z.}~\bibnamefont {Fang}}, \bibinfo {author} {\bibfnamefont {S.-C.}\
  \bibnamefont {Zhang}}, \bibinfo {author} {\bibfnamefont {K.}~\bibnamefont
  {He}}, \bibinfo {author} {\bibfnamefont {Y.}~\bibnamefont {Wang}}, \bibinfo
  {author} {\bibfnamefont {L.}~\bibnamefont {Lu}}, \bibinfo {author}
  {\bibfnamefont {X.-C.}\ \bibnamefont {Ma}},\ and\ \bibinfo {author}
  {\bibfnamefont {Q.-K.}\ \bibnamefont {Xue}},\ }\href
  {https://doi.org/10.1126/science.1234414} {\bibfield  {journal} {\bibinfo
  {journal} {Science}\ }\textbf {\bibinfo {volume} {340}},\ \bibinfo {pages}
  {167} (\bibinfo {year} {2013})}\BibitemShut {NoStop}%
\bibitem [{\citenamefont {Serlin}\ \emph {et~al.}(2020)\citenamefont {Serlin},
  \citenamefont {Tschirhart}, \citenamefont {Polshyn}, \citenamefont {Zhang},
  \citenamefont {Zhu}, \citenamefont {Watanabe}, \citenamefont {Taniguchi},
  \citenamefont {Balents},\ and\ \citenamefont
  {Young}}]{serlin_intrinsic_2020}%
  \BibitemOpen
  \bibfield  {author} {\bibinfo {author} {\bibfnamefont {M.}~\bibnamefont
  {Serlin}}, \bibinfo {author} {\bibfnamefont {C.~L.}\ \bibnamefont
  {Tschirhart}}, \bibinfo {author} {\bibfnamefont {H.}~\bibnamefont {Polshyn}},
  \bibinfo {author} {\bibfnamefont {Y.}~\bibnamefont {Zhang}}, \bibinfo
  {author} {\bibfnamefont {J.}~\bibnamefont {Zhu}}, \bibinfo {author}
  {\bibfnamefont {K.}~\bibnamefont {Watanabe}}, \bibinfo {author}
  {\bibfnamefont {T.}~\bibnamefont {Taniguchi}}, \bibinfo {author}
  {\bibfnamefont {L.}~\bibnamefont {Balents}},\ and\ \bibinfo {author}
  {\bibfnamefont {A.~F.}\ \bibnamefont {Young}},\ }\href
  {https://doi.org/10.1126/science.aay5533} {\bibfield  {journal} {\bibinfo
  {journal} {Science}\ }\textbf {\bibinfo {volume} {367}},\ \bibinfo {pages}
  {900} (\bibinfo {year} {2020})}\BibitemShut {NoStop}%
\bibitem [{\citenamefont {Deng}\ \emph {et~al.}(2020)\citenamefont {Deng},
  \citenamefont {Yu}, \citenamefont {Shi}, \citenamefont {Guo}, \citenamefont
  {Xu}, \citenamefont {Wang}, \citenamefont {Chen},\ and\ \citenamefont
  {Zhang}}]{deng_quantum_2020}%
  \BibitemOpen
  \bibfield  {author} {\bibinfo {author} {\bibfnamefont {Y.}~\bibnamefont
  {Deng}}, \bibinfo {author} {\bibfnamefont {Y.}~\bibnamefont {Yu}}, \bibinfo
  {author} {\bibfnamefont {M.~Z.}\ \bibnamefont {Shi}}, \bibinfo {author}
  {\bibfnamefont {Z.}~\bibnamefont {Guo}}, \bibinfo {author} {\bibfnamefont
  {Z.}~\bibnamefont {Xu}}, \bibinfo {author} {\bibfnamefont {J.}~\bibnamefont
  {Wang}}, \bibinfo {author} {\bibfnamefont {X.~H.}\ \bibnamefont {Chen}},\
  and\ \bibinfo {author} {\bibfnamefont {Y.}~\bibnamefont {Zhang}},\ }\href
  {https://doi.org/10.1126/science.aax8156} {\bibfield  {journal} {\bibinfo
  {journal} {Science}\ }\textbf {\bibinfo {volume} {367}},\ \bibinfo {pages}
  {895} (\bibinfo {year} {2020})}\BibitemShut {NoStop}%
\bibitem [{\citenamefont {Wang}\ \emph {et~al.}(2009)\citenamefont {Wang},
  \citenamefont {Chong}, \citenamefont {Joannopoulos},\ and\ \citenamefont
  {Solja{\v c}i{\'c}}}]{wang_observation_2009}%
  \BibitemOpen
  \bibfield  {author} {\bibinfo {author} {\bibfnamefont {Z.}~\bibnamefont
  {Wang}}, \bibinfo {author} {\bibfnamefont {Y.}~\bibnamefont {Chong}},
  \bibinfo {author} {\bibfnamefont {J.~D.}\ \bibnamefont {Joannopoulos}},\ and\
  \bibinfo {author} {\bibfnamefont {M.}~\bibnamefont {Solja{\v c}i{\'c}}},\
  }\href {https://doi.org/10.1038/nature08293} {\bibfield  {journal} {\bibinfo
  {journal} {Nature}\ }\textbf {\bibinfo {volume} {461}},\ \bibinfo {pages}
  {772} (\bibinfo {year} {2009})}\BibitemShut {NoStop}%
\bibitem [{\citenamefont {Jotzu}\ \emph {et~al.}(2014)\citenamefont {Jotzu},
  \citenamefont {Messer}, \citenamefont {Desbuquois}, \citenamefont {Lebrat},
  \citenamefont {Uehlinger}, \citenamefont {Greif},\ and\ \citenamefont
  {Esslinger}}]{jotzu_experimental_2014}%
  \BibitemOpen
  \bibfield  {author} {\bibinfo {author} {\bibfnamefont {G.}~\bibnamefont
  {Jotzu}}, \bibinfo {author} {\bibfnamefont {M.}~\bibnamefont {Messer}},
  \bibinfo {author} {\bibfnamefont {R.}~\bibnamefont {Desbuquois}}, \bibinfo
  {author} {\bibfnamefont {M.}~\bibnamefont {Lebrat}}, \bibinfo {author}
  {\bibfnamefont {T.}~\bibnamefont {Uehlinger}}, \bibinfo {author}
  {\bibfnamefont {D.}~\bibnamefont {Greif}},\ and\ \bibinfo {author}
  {\bibfnamefont {T.}~\bibnamefont {Esslinger}},\ }\href
  {https://doi.org/10.1038/nature13915} {\bibfield  {journal} {\bibinfo
  {journal} {Nature}\ }\textbf {\bibinfo {volume} {515}},\ \bibinfo {pages}
  {237} (\bibinfo {year} {2014})}\BibitemShut {NoStop}%
\bibitem [{\citenamefont {Klembt}\ \emph {et~al.}(2018)\citenamefont {Klembt},
  \citenamefont {Harder}, \citenamefont {Egorov}, \citenamefont {Winkler},
  \citenamefont {Ge}, \citenamefont {Bandres}, \citenamefont {Emmerling},
  \citenamefont {Worschech}, \citenamefont {Liew}, \citenamefont {Segev},
  \citenamefont {Schneider},\ and\ \citenamefont
  {H{\"o}fling}}]{klembt_exciton-polariton_2018}%
  \BibitemOpen
  \bibfield  {author} {\bibinfo {author} {\bibfnamefont {S.}~\bibnamefont
  {Klembt}}, \bibinfo {author} {\bibfnamefont {T.~H.}\ \bibnamefont {Harder}},
  \bibinfo {author} {\bibfnamefont {O.~A.}\ \bibnamefont {Egorov}}, \bibinfo
  {author} {\bibfnamefont {K.}~\bibnamefont {Winkler}}, \bibinfo {author}
  {\bibfnamefont {R.}~\bibnamefont {Ge}}, \bibinfo {author} {\bibfnamefont
  {M.~A.}\ \bibnamefont {Bandres}}, \bibinfo {author} {\bibfnamefont
  {M.}~\bibnamefont {Emmerling}}, \bibinfo {author} {\bibfnamefont
  {L.}~\bibnamefont {Worschech}}, \bibinfo {author} {\bibfnamefont {T.~C.~H.}\
  \bibnamefont {Liew}}, \bibinfo {author} {\bibfnamefont {M.}~\bibnamefont
  {Segev}}, \bibinfo {author} {\bibfnamefont {C.}~\bibnamefont {Schneider}},\
  and\ \bibinfo {author} {\bibfnamefont {S.}~\bibnamefont {H{\"o}fling}},\
  }\href {https://doi.org/10.1038/s41586-018-0601-5} {\bibfield  {journal}
  {\bibinfo  {journal} {Nature}\ }\textbf {\bibinfo {volume} {562}},\ \bibinfo
  {pages} {552} (\bibinfo {year} {2018})}\BibitemShut {NoStop}%
\bibitem [{\citenamefont {Fu}\ \emph {et~al.}(2007)\citenamefont {Fu},
  \citenamefont {Kane},\ and\ \citenamefont {Mele}}]{fu_topological_2007}%
  \BibitemOpen
  \bibfield  {author} {\bibinfo {author} {\bibfnamefont {L.}~\bibnamefont
  {Fu}}, \bibinfo {author} {\bibfnamefont {C.~L.}\ \bibnamefont {Kane}},\ and\
  \bibinfo {author} {\bibfnamefont {E.~J.}\ \bibnamefont {Mele}},\ }\href
  {https://doi.org/10.1103/PhysRevLett.98.106803} {\bibfield  {journal}
  {\bibinfo  {journal} {Phys. Rev. Lett.}\ }\textbf {\bibinfo {volume} {98}},\
  \bibinfo {pages} {106803} (\bibinfo {year} {2007})}\BibitemShut {NoStop}%
\bibitem [{\citenamefont {Lu}\ \emph {et~al.}(2021)\citenamefont {Lu},
  \citenamefont {Sun}, \citenamefont {Kumar}, \citenamefont {Wang},
  \citenamefont {Gu}, \citenamefont {Zeng}, \citenamefont {Hao}, \citenamefont
  {Li}, \citenamefont {Shao}, \citenamefont {Ma}, \citenamefont {Hao},
  \citenamefont {Zhang}, \citenamefont {Mansuer}, \citenamefont {Mei},
  \citenamefont {Zhao}, \citenamefont {Liu}, \citenamefont {Deng},
  \citenamefont {Huang}, \citenamefont {Shen}, \citenamefont {Shimada},
  \citenamefont {Schwier}, \citenamefont {Liu}, \citenamefont {Liu},\ and\
  \citenamefont {Chen}}]{lu_half-magnetic_2021}%
  \BibitemOpen
  \bibfield  {author} {\bibinfo {author} {\bibfnamefont {R.}~\bibnamefont
  {Lu}}, \bibinfo {author} {\bibfnamefont {H.}~\bibnamefont {Sun}}, \bibinfo
  {author} {\bibfnamefont {S.}~\bibnamefont {Kumar}}, \bibinfo {author}
  {\bibfnamefont {Y.}~\bibnamefont {Wang}}, \bibinfo {author} {\bibfnamefont
  {M.}~\bibnamefont {Gu}}, \bibinfo {author} {\bibfnamefont {M.}~\bibnamefont
  {Zeng}}, \bibinfo {author} {\bibfnamefont {Y.-J.}\ \bibnamefont {Hao}},
  \bibinfo {author} {\bibfnamefont {J.}~\bibnamefont {Li}}, \bibinfo {author}
  {\bibfnamefont {J.}~\bibnamefont {Shao}}, \bibinfo {author} {\bibfnamefont
  {X.-M.}\ \bibnamefont {Ma}}, \bibinfo {author} {\bibfnamefont
  {Z.}~\bibnamefont {Hao}}, \bibinfo {author} {\bibfnamefont {K.}~\bibnamefont
  {Zhang}}, \bibinfo {author} {\bibfnamefont {W.}~\bibnamefont {Mansuer}},
  \bibinfo {author} {\bibfnamefont {J.}~\bibnamefont {Mei}}, \bibinfo {author}
  {\bibfnamefont {Y.}~\bibnamefont {Zhao}}, \bibinfo {author} {\bibfnamefont
  {C.}~\bibnamefont {Liu}}, \bibinfo {author} {\bibfnamefont {K.}~\bibnamefont
  {Deng}}, \bibinfo {author} {\bibfnamefont {W.}~\bibnamefont {Huang}},
  \bibinfo {author} {\bibfnamefont {B.}~\bibnamefont {Shen}}, \bibinfo {author}
  {\bibfnamefont {K.}~\bibnamefont {Shimada}}, \bibinfo {author} {\bibfnamefont
  {E.~F.}\ \bibnamefont {Schwier}}, \bibinfo {author} {\bibfnamefont
  {C.}~\bibnamefont {Liu}}, \bibinfo {author} {\bibfnamefont {Q.}~\bibnamefont
  {Liu}},\ and\ \bibinfo {author} {\bibfnamefont {C.}~\bibnamefont {Chen}},\
  }\href {https://doi.org/10.1103/PhysRevX.11.011039} {\bibfield  {journal}
  {\bibinfo  {journal} {Phys. Rev. X}\ }\textbf {\bibinfo {volume} {11}},\
  \bibinfo {pages} {011039} (\bibinfo {year} {2021})}\BibitemShut {NoStop}%
\bibitem [{\citenamefont {Mogi}\ \emph {et~al.}(2022)\citenamefont {Mogi},
  \citenamefont {Okamura}, \citenamefont {Kawamura}, \citenamefont {Yoshimi},
  \citenamefont {Yasuda}, \citenamefont {Tsukazaki}, \citenamefont {Takahashi},
  \citenamefont {Morimoto}, \citenamefont {Nagaosa}, \citenamefont {Kawasaki},
  \citenamefont {Takahashi},\ and\ \citenamefont
  {Tokura}}]{mogi_experimental_2022}%
  \BibitemOpen
  \bibfield  {author} {\bibinfo {author} {\bibfnamefont {M.}~\bibnamefont
  {Mogi}}, \bibinfo {author} {\bibfnamefont {Y.}~\bibnamefont {Okamura}},
  \bibinfo {author} {\bibfnamefont {M.}~\bibnamefont {Kawamura}}, \bibinfo
  {author} {\bibfnamefont {R.}~\bibnamefont {Yoshimi}}, \bibinfo {author}
  {\bibfnamefont {K.}~\bibnamefont {Yasuda}}, \bibinfo {author} {\bibfnamefont
  {A.}~\bibnamefont {Tsukazaki}}, \bibinfo {author} {\bibfnamefont {K.~S.}\
  \bibnamefont {Takahashi}}, \bibinfo {author} {\bibfnamefont {T.}~\bibnamefont
  {Morimoto}}, \bibinfo {author} {\bibfnamefont {N.}~\bibnamefont {Nagaosa}},
  \bibinfo {author} {\bibfnamefont {M.}~\bibnamefont {Kawasaki}}, \bibinfo
  {author} {\bibfnamefont {Y.}~\bibnamefont {Takahashi}},\ and\ \bibinfo
  {author} {\bibfnamefont {Y.}~\bibnamefont {Tokura}},\ }\href
  {https://doi.org/10.1038/s41567-021-01490-y} {\bibfield  {journal} {\bibinfo
  {journal} {Nat. Phys.}\ }\textbf {\bibinfo {volume} {18}},\ \bibinfo {pages}
  {390} (\bibinfo {year} {2022})}\BibitemShut {NoStop}%
\bibitem [{\citenamefont {Leykam}\ \emph {et~al.}(2016)\citenamefont {Leykam},
  \citenamefont {Rechtsman},\ and\ \citenamefont
  {Chong}}]{leykam_anomalous_2016}%
  \BibitemOpen
  \bibfield  {author} {\bibinfo {author} {\bibfnamefont {D.}~\bibnamefont
  {Leykam}}, \bibinfo {author} {\bibfnamefont {M.~C.}\ \bibnamefont
  {Rechtsman}},\ and\ \bibinfo {author} {\bibfnamefont {Y.~D.}\ \bibnamefont
  {Chong}},\ }\href {https://doi.org/10.1103/PhysRevLett.117.013902} {\bibfield
   {journal} {\bibinfo  {journal} {Phys. Rev. Lett.}\ }\textbf {\bibinfo
  {volume} {117}},\ \bibinfo {pages} {013902} (\bibinfo {year}
  {2016})}\BibitemShut {NoStop}%
\bibitem [{\citenamefont {Liu}\ \emph {et~al.}(2020)\citenamefont {Liu},
  \citenamefont {Zhou}, \citenamefont {Yang}, \citenamefont {Xue},
  \citenamefont {Ren}, \citenamefont {Lin}, \citenamefont {Sun}, \citenamefont
  {Bi}, \citenamefont {Chong},\ and\ \citenamefont
  {Zhang}}]{liu_observation_2020}%
  \BibitemOpen
  \bibfield  {author} {\bibinfo {author} {\bibfnamefont {G.-G.}\ \bibnamefont
  {Liu}}, \bibinfo {author} {\bibfnamefont {P.}~\bibnamefont {Zhou}}, \bibinfo
  {author} {\bibfnamefont {Y.}~\bibnamefont {Yang}}, \bibinfo {author}
  {\bibfnamefont {H.}~\bibnamefont {Xue}}, \bibinfo {author} {\bibfnamefont
  {X.}~\bibnamefont {Ren}}, \bibinfo {author} {\bibfnamefont {X.}~\bibnamefont
  {Lin}}, \bibinfo {author} {\bibfnamefont {H.-x.}\ \bibnamefont {Sun}},
  \bibinfo {author} {\bibfnamefont {L.}~\bibnamefont {Bi}}, \bibinfo {author}
  {\bibfnamefont {Y.}~\bibnamefont {Chong}},\ and\ \bibinfo {author}
  {\bibfnamefont {B.}~\bibnamefont {Zhang}},\ }\href
  {https://doi.org/10.1038/s41467-020-15801-z} {\bibfield  {journal} {\bibinfo
  {journal} {Nat Commun}\ }\textbf {\bibinfo {volume} {11}},\ \bibinfo {pages}
  {1873} (\bibinfo {year} {2020})}\BibitemShut {NoStop}%
\bibitem [{\citenamefont {Zhong}\ \emph {et~al.}(2024)\citenamefont {Zhong},
  \citenamefont {Kartashov}, \citenamefont {Li}, \citenamefont {Li},\ and\
  \citenamefont {Zhang}}]{zhong_topological_2024}%
  \BibitemOpen
  \bibfield  {author} {\bibinfo {author} {\bibfnamefont {H.}~\bibnamefont
  {Zhong}}, \bibinfo {author} {\bibfnamefont {Y.~V.}\ \bibnamefont
  {Kartashov}}, \bibinfo {author} {\bibfnamefont {Y.}~\bibnamefont {Li}},
  \bibinfo {author} {\bibfnamefont {M.}~\bibnamefont {Li}},\ and\ \bibinfo
  {author} {\bibfnamefont {Y.}~\bibnamefont {Zhang}},\ }\href
  {https://doi.org/10.1364/PRJ.524824} {\bibfield  {journal} {\bibinfo
  {journal} {Photon. Res., PRJ}\ }\textbf {\bibinfo {volume} {12}},\ \bibinfo
  {pages} {2078} (\bibinfo {year} {2024})}\BibitemShut {NoStop}%
\bibitem [{\citenamefont {Yuan}\ \emph {et~al.}(2025)\citenamefont {Yuan},
  \citenamefont {Yi}, \citenamefont {Guo}, \citenamefont {Cheng}, \citenamefont
  {Jiao}, \citenamefont {Zhang}, \citenamefont {Chen},\ and\ \citenamefont
  {Pan}}]{yuan_observation_2025}%
  \BibitemOpen
  \bibfield  {author} {\bibinfo {author} {\bibfnamefont {H.}~\bibnamefont
  {Yuan}}, \bibinfo {author} {\bibfnamefont {C.-R.}\ \bibnamefont {Yi}},
  \bibinfo {author} {\bibfnamefont {J.-Y.}\ \bibnamefont {Guo}}, \bibinfo
  {author} {\bibfnamefont {X.-C.}\ \bibnamefont {Cheng}}, \bibinfo {author}
  {\bibfnamefont {R.-H.}\ \bibnamefont {Jiao}}, \bibinfo {author}
  {\bibfnamefont {J.}~\bibnamefont {Zhang}}, \bibinfo {author} {\bibfnamefont
  {S.}~\bibnamefont {Chen}},\ and\ \bibinfo {author} {\bibfnamefont {J.-W.}\
  \bibnamefont {Pan}},\ }\href {https://doi.org/10.1103/5j5s-946k} {\bibfield
  {journal} {\bibinfo  {journal} {Phys. Rev. Lett.}\ }\textbf {\bibinfo
  {volume} {135}},\ \bibinfo {pages} {063403} (\bibinfo {year}
  {2025})}\BibitemShut {NoStop}%
\bibitem [{\citenamefont {Celi}\ \emph {et~al.}(2014)\citenamefont {Celi},
  \citenamefont {Massignan}, \citenamefont {Ruseckas}, \citenamefont {Goldman},
  \citenamefont {Spielman}, \citenamefont {Juzeli{\=u}nas},\ and\ \citenamefont
  {Lewenstein}}]{celi_synthetic_2014}%
  \BibitemOpen
  \bibfield  {author} {\bibinfo {author} {\bibfnamefont {A.}~\bibnamefont
  {Celi}}, \bibinfo {author} {\bibfnamefont {P.}~\bibnamefont {Massignan}},
  \bibinfo {author} {\bibfnamefont {J.}~\bibnamefont {Ruseckas}}, \bibinfo
  {author} {\bibfnamefont {N.}~\bibnamefont {Goldman}}, \bibinfo {author}
  {\bibfnamefont {I.~B.}\ \bibnamefont {Spielman}}, \bibinfo {author}
  {\bibfnamefont {G.}~\bibnamefont {Juzeli{\=u}nas}},\ and\ \bibinfo {author}
  {\bibfnamefont {M.}~\bibnamefont {Lewenstein}},\ }\href
  {https://doi.org/10.1103/PhysRevLett.112.043001} {\bibfield  {journal}
  {\bibinfo  {journal} {Phys. Rev. Lett.}\ }\textbf {\bibinfo {volume} {112}},\
  \bibinfo {pages} {043001} (\bibinfo {year} {2014})}\BibitemShut {NoStop}%
\bibitem [{\citenamefont {Mancini}\ \emph {et~al.}(2015)\citenamefont
  {Mancini}, \citenamefont {Pagano}, \citenamefont {Cappellini}, \citenamefont
  {Livi}, \citenamefont {Rider}, \citenamefont {Catani}, \citenamefont {Sias},
  \citenamefont {Zoller}, \citenamefont {Inguscio}, \citenamefont {Dalmonte},\
  and\ \citenamefont {Fallani}}]{mancini_observation_2015}%
  \BibitemOpen
  \bibfield  {author} {\bibinfo {author} {\bibfnamefont {M.}~\bibnamefont
  {Mancini}}, \bibinfo {author} {\bibfnamefont {G.}~\bibnamefont {Pagano}},
  \bibinfo {author} {\bibfnamefont {G.}~\bibnamefont {Cappellini}}, \bibinfo
  {author} {\bibfnamefont {L.}~\bibnamefont {Livi}}, \bibinfo {author}
  {\bibfnamefont {M.}~\bibnamefont {Rider}}, \bibinfo {author} {\bibfnamefont
  {J.}~\bibnamefont {Catani}}, \bibinfo {author} {\bibfnamefont
  {C.}~\bibnamefont {Sias}}, \bibinfo {author} {\bibfnamefont {P.}~\bibnamefont
  {Zoller}}, \bibinfo {author} {\bibfnamefont {M.}~\bibnamefont {Inguscio}},
  \bibinfo {author} {\bibfnamefont {M.}~\bibnamefont {Dalmonte}},\ and\
  \bibinfo {author} {\bibfnamefont {L.}~\bibnamefont {Fallani}},\ }\href
  {https://doi.org/10.1126/science.aaa8736} {\bibfield  {journal} {\bibinfo
  {journal} {Science}\ }\textbf {\bibinfo {volume} {349}},\ \bibinfo {pages}
  {1510} (\bibinfo {year} {2015})}\BibitemShut {NoStop}%
\bibitem [{\citenamefont {Stuhl}\ \emph {et~al.}(2015)\citenamefont {Stuhl},
  \citenamefont {Lu}, \citenamefont {Aycock}, \citenamefont {Genkina},\ and\
  \citenamefont {Spielman}}]{stuhl_visualizing_2015}%
  \BibitemOpen
  \bibfield  {author} {\bibinfo {author} {\bibfnamefont {B.~K.}\ \bibnamefont
  {Stuhl}}, \bibinfo {author} {\bibfnamefont {H.-I.}\ \bibnamefont {Lu}},
  \bibinfo {author} {\bibfnamefont {L.~M.}\ \bibnamefont {Aycock}}, \bibinfo
  {author} {\bibfnamefont {D.}~\bibnamefont {Genkina}},\ and\ \bibinfo {author}
  {\bibfnamefont {I.~B.}\ \bibnamefont {Spielman}},\ }\href
  {https://doi.org/10.1126/science.aaa8515} {\bibfield  {journal} {\bibinfo
  {journal} {Science}\ }\textbf {\bibinfo {volume} {349}},\ \bibinfo {pages}
  {1514} (\bibinfo {year} {2015})}\BibitemShut {NoStop}%
\bibitem [{\citenamefont {Chalopin}\ \emph {et~al.}(2020)\citenamefont
  {Chalopin}, \citenamefont {Satoor}, \citenamefont {Evrard}, \citenamefont
  {Makhalov}, \citenamefont {Dalibard}, \citenamefont {Lopes},\ and\
  \citenamefont {Nascimbene}}]{chalopin_probing_2020}%
  \BibitemOpen
  \bibfield  {author} {\bibinfo {author} {\bibfnamefont {T.}~\bibnamefont
  {Chalopin}}, \bibinfo {author} {\bibfnamefont {T.}~\bibnamefont {Satoor}},
  \bibinfo {author} {\bibfnamefont {A.}~\bibnamefont {Evrard}}, \bibinfo
  {author} {\bibfnamefont {V.}~\bibnamefont {Makhalov}}, \bibinfo {author}
  {\bibfnamefont {J.}~\bibnamefont {Dalibard}}, \bibinfo {author}
  {\bibfnamefont {R.}~\bibnamefont {Lopes}},\ and\ \bibinfo {author}
  {\bibfnamefont {S.}~\bibnamefont {Nascimbene}},\ }\href
  {https://doi.org/10.1038/s41567-020-0942-5} {\bibfield  {journal} {\bibinfo
  {journal} {Nat. Phys.}\ }\textbf {\bibinfo {volume} {16}},\ \bibinfo {pages}
  {1017} (\bibinfo {year} {2020})}\BibitemShut {NoStop}%
\bibitem [{\citenamefont {Kane}\ \emph {et~al.}(2002)\citenamefont {Kane},
  \citenamefont {Mukhopadhyay},\ and\ \citenamefont
  {Lubensky}}]{kane_fractional_2002}%
  \BibitemOpen
  \bibfield  {author} {\bibinfo {author} {\bibfnamefont {C.~L.}\ \bibnamefont
  {Kane}}, \bibinfo {author} {\bibfnamefont {R.}~\bibnamefont {Mukhopadhyay}},\
  and\ \bibinfo {author} {\bibfnamefont {T.~C.}\ \bibnamefont {Lubensky}},\
  }\href {https://doi.org/10.1103/PhysRevLett.88.036401} {\bibfield  {journal}
  {\bibinfo  {journal} {Phys. Rev. Lett.}\ }\textbf {\bibinfo {volume} {88}},\
  \bibinfo {pages} {036401} (\bibinfo {year} {2002})}\BibitemShut {NoStop}%
\bibitem [{\citenamefont {Thouless}(1983)}]{thouless_quantization_1983}%
  \BibitemOpen
  \bibfield  {author} {\bibinfo {author} {\bibfnamefont {D.~J.}\ \bibnamefont
  {Thouless}},\ }\href {https://doi.org/10.1103/PhysRevB.27.6083} {\bibfield
  {journal} {\bibinfo  {journal} {Phys. Rev. B}\ }\textbf {\bibinfo {volume}
  {27}},\ \bibinfo {pages} {6083} (\bibinfo {year} {1983})}\BibitemShut
  {NoStop}%
\bibitem [{\citenamefont {Szumniak}\ \emph {et~al.}(2016)\citenamefont
  {Szumniak}, \citenamefont {Klinovaja},\ and\ \citenamefont
  {Loss}}]{szumniak_chiral_2016}%
  \BibitemOpen
  \bibfield  {author} {\bibinfo {author} {\bibfnamefont {P.}~\bibnamefont
  {Szumniak}}, \bibinfo {author} {\bibfnamefont {J.}~\bibnamefont
  {Klinovaja}},\ and\ \bibinfo {author} {\bibfnamefont {D.}~\bibnamefont
  {Loss}},\ }\href {https://doi.org/10.1103/PhysRevB.93.245308} {\bibfield
  {journal} {\bibinfo  {journal} {Phys. Rev. B}\ }\textbf {\bibinfo {volume}
  {93}},\ \bibinfo {pages} {245308} (\bibinfo {year} {2016})}\BibitemShut
  {NoStop}%
\bibitem [{\citenamefont {Resta}\ and\ \citenamefont
  {Vanderbilt}(2007)}]{resta_theory_2007}%
  \BibitemOpen
  \bibfield  {author} {\bibinfo {author} {\bibfnamefont {R.}~\bibnamefont
  {Resta}}\ and\ \bibinfo {author} {\bibfnamefont {D.}~\bibnamefont
  {Vanderbilt}},\ }in\ \href {https://doi.org/10.1007/978-3-540-34591-6_2}
  {\emph {\bibinfo {booktitle} {Physics of {{Ferroelectrics}}: {{A Modern
  Perspective}}}}}\ (\bibinfo  {publisher} {Springer},\ \bibinfo {address}
  {Berlin, Heidelberg},\ \bibinfo {year} {2007})\ pp.\ \bibinfo {pages}
  {31--68}\BibitemShut {NoStop}%
\bibitem [{\citenamefont {Semenoff}(1984)}]{semenoff_condensed-matter_1984}%
  \BibitemOpen
  \bibfield  {author} {\bibinfo {author} {\bibfnamefont {G.~W.}\ \bibnamefont
  {Semenoff}},\ }\href {https://doi.org/10.1103/PhysRevLett.53.2449} {\bibfield
   {journal} {\bibinfo  {journal} {Phys. Rev. Lett.}\ }\textbf {\bibinfo
  {volume} {53}},\ \bibinfo {pages} {2449} (\bibinfo {year}
  {1984})}\BibitemShut {NoStop}%
\bibitem [{\citenamefont {Fu}\ \emph {et~al.}(2022)\citenamefont {Fu},
  \citenamefont {Zou}, \citenamefont {Hu}, \citenamefont {Wang},\ and\
  \citenamefont {Shen}}]{fu_quantum_2022}%
  \BibitemOpen
  \bibfield  {author} {\bibinfo {author} {\bibfnamefont {B.}~\bibnamefont
  {Fu}}, \bibinfo {author} {\bibfnamefont {J.-Y.}\ \bibnamefont {Zou}},
  \bibinfo {author} {\bibfnamefont {Z.-A.}\ \bibnamefont {Hu}}, \bibinfo
  {author} {\bibfnamefont {H.-W.}\ \bibnamefont {Wang}},\ and\ \bibinfo
  {author} {\bibfnamefont {S.-Q.}\ \bibnamefont {Shen}},\ }\href
  {https://doi.org/10.1038/s41535-022-00503-0} {\bibfield  {journal} {\bibinfo
  {journal} {npj Quantum Mater.}\ }\textbf {\bibinfo {volume} {7}},\ \bibinfo
  {pages} {94} (\bibinfo {year} {2022})}\BibitemShut {NoStop}%
\bibitem [{\citenamefont {Bernevig}\ and\ \citenamefont
  {Hughes}(2013)}]{bernevig_topological_2013}%
  \BibitemOpen
  \bibfield  {author} {\bibinfo {author} {\bibfnamefont {B.~A.}\ \bibnamefont
  {Bernevig}}\ and\ \bibinfo {author} {\bibfnamefont {T.~L.}\ \bibnamefont
  {Hughes}},\ }\href@noop {} {\emph {\bibinfo {title} {{Topological Insulators
  and Topological Superconductors}}}}\ (\bibinfo  {publisher} {Princeton
  University Press},\ \bibinfo {address} {Princeton},\ \bibinfo {year}
  {2013})\BibitemShut {NoStop}%
\bibitem [{\citenamefont {Carruthers}\ and\ \citenamefont
  {Nieto}(1968)}]{carruthers_phase_1968}%
  \BibitemOpen
  \bibfield  {author} {\bibinfo {author} {\bibfnamefont {P.}~\bibnamefont
  {Carruthers}}\ and\ \bibinfo {author} {\bibfnamefont {M.~M.}\ \bibnamefont
  {Nieto}},\ }\href {https://doi.org/10.1103/RevModPhys.40.411} {\bibfield
  {journal} {\bibinfo  {journal} {Rev. Mod. Phys.}\ }\textbf {\bibinfo {volume}
  {40}},\ \bibinfo {pages} {411} (\bibinfo {year} {1968})}\BibitemShut
  {NoStop}%
\bibitem [{\citenamefont {Aidelsburger}\ \emph {et~al.}(2015)\citenamefont
  {Aidelsburger}, \citenamefont {Lohse}, \citenamefont {Schweizer},
  \citenamefont {Atala}, \citenamefont {Barreiro}, \citenamefont {Nascimbene},
  \citenamefont {Cooper}, \citenamefont {Bloch},\ and\ \citenamefont
  {Goldman}}]{aidelsburger_measuring_2015}%
  \BibitemOpen
  \bibfield  {author} {\bibinfo {author} {\bibfnamefont {M.}~\bibnamefont
  {Aidelsburger}}, \bibinfo {author} {\bibfnamefont {M.}~\bibnamefont {Lohse}},
  \bibinfo {author} {\bibfnamefont {C.}~\bibnamefont {Schweizer}}, \bibinfo
  {author} {\bibfnamefont {M.}~\bibnamefont {Atala}}, \bibinfo {author}
  {\bibfnamefont {J.~T.}\ \bibnamefont {Barreiro}}, \bibinfo {author}
  {\bibfnamefont {S.}~\bibnamefont {Nascimbene}}, \bibinfo {author}
  {\bibfnamefont {N.~R.}\ \bibnamefont {Cooper}}, \bibinfo {author}
  {\bibfnamefont {I.}~\bibnamefont {Bloch}},\ and\ \bibinfo {author}
  {\bibfnamefont {N.}~\bibnamefont {Goldman}},\ }\href
  {https://doi.org/doi.org/10.1038/nphys3171} {\bibfield  {journal} {\bibinfo
  {journal} {Nat. Phys.}\ }\textbf {\bibinfo {volume} {11}},\ \bibinfo {pages}
  {162} (\bibinfo {year} {2015})}\BibitemShut {NoStop}%
\bibitem [{\citenamefont {Asteria}\ \emph {et~al.}(2019)\citenamefont
  {Asteria}, \citenamefont {Tran}, \citenamefont {Ozawa}, \citenamefont
  {Tarnowski}, \citenamefont {Rem}, \citenamefont {Fl{\"a}schner},
  \citenamefont {Sengstock}, \citenamefont {Goldman},\ and\ \citenamefont
  {Weitenberg}}]{asteria_measuring_2019}%
  \BibitemOpen
  \bibfield  {author} {\bibinfo {author} {\bibfnamefont {L.}~\bibnamefont
  {Asteria}}, \bibinfo {author} {\bibfnamefont {D.~T.}\ \bibnamefont {Tran}},
  \bibinfo {author} {\bibfnamefont {T.}~\bibnamefont {Ozawa}}, \bibinfo
  {author} {\bibfnamefont {M.}~\bibnamefont {Tarnowski}}, \bibinfo {author}
  {\bibfnamefont {B.~S.}\ \bibnamefont {Rem}}, \bibinfo {author} {\bibfnamefont
  {N.}~\bibnamefont {Fl{\"a}schner}}, \bibinfo {author} {\bibfnamefont
  {K.}~\bibnamefont {Sengstock}}, \bibinfo {author} {\bibfnamefont
  {N.}~\bibnamefont {Goldman}},\ and\ \bibinfo {author} {\bibfnamefont
  {C.}~\bibnamefont {Weitenberg}},\ }\href
  {https://doi.org/10.1038/s41567-019-0417-8} {\bibfield  {journal} {\bibinfo
  {journal} {Nature Phys.}\ }\textbf {\bibinfo {volume} {15}},\ \bibinfo
  {pages} {449} (\bibinfo {year} {2019})}\BibitemShut {NoStop}%
\bibitem [{\citenamefont {Ben~Dahan}\ \emph {et~al.}(1996)\citenamefont
  {Ben~Dahan}, \citenamefont {Peik}, \citenamefont {Reichel}, \citenamefont
  {Castin},\ and\ \citenamefont {Salomon}}]{ben_dahan_bloch_1996}%
  \BibitemOpen
  \bibfield  {author} {\bibinfo {author} {\bibfnamefont {M.}~\bibnamefont
  {Ben~Dahan}}, \bibinfo {author} {\bibfnamefont {E.}~\bibnamefont {Peik}},
  \bibinfo {author} {\bibfnamefont {J.}~\bibnamefont {Reichel}}, \bibinfo
  {author} {\bibfnamefont {Y.}~\bibnamefont {Castin}},\ and\ \bibinfo {author}
  {\bibfnamefont {C.}~\bibnamefont {Salomon}},\ }\href
  {https://doi.org/10.1103/PhysRevLett.76.4508} {\bibfield  {journal} {\bibinfo
   {journal} {Phys. Rev. Lett.}\ }\textbf {\bibinfo {volume} {76}},\ \bibinfo
  {pages} {4508} (\bibinfo {year} {1996})}\BibitemShut {NoStop}%
\bibitem [{\citenamefont {Bianco}\ and\ \citenamefont
  {Resta}(2011)}]{bianco_mapping_2011}%
  \BibitemOpen
  \bibfield  {author} {\bibinfo {author} {\bibfnamefont {R.}~\bibnamefont
  {Bianco}}\ and\ \bibinfo {author} {\bibfnamefont {R.}~\bibnamefont {Resta}},\
  }\href {https://doi.org/10.1103/PhysRevB.84.241106} {\bibfield  {journal}
  {\bibinfo  {journal} {Phys. Rev. B}\ }\textbf {\bibinfo {volume} {84}},\
  \bibinfo {pages} {241106} (\bibinfo {year} {2011})}\BibitemShut {NoStop}%
\bibitem [{\citenamefont {Wen}(1990)}]{wen_chiral_1990}%
  \BibitemOpen
  \bibfield  {author} {\bibinfo {author} {\bibfnamefont {X.~G.}\ \bibnamefont
  {Wen}},\ }\href {https://doi.org/10.1103/PhysRevB.41.12838} {\bibfield
  {journal} {\bibinfo  {journal} {Phys. Rev. B}\ }\textbf {\bibinfo {volume}
  {41}},\ \bibinfo {pages} {12838} (\bibinfo {year} {1990})}\BibitemShut
  {NoStop}%
\bibitem [{\citenamefont {Chen}\ \emph {et~al.}(2017)\citenamefont {Chen},
  \citenamefont {Legner}, \citenamefont {R{\"u}egg},\ and\ \citenamefont
  {Sigrist}}]{chen_correlation_2017}%
  \BibitemOpen
  \bibfield  {author} {\bibinfo {author} {\bibfnamefont {W.}~\bibnamefont
  {Chen}}, \bibinfo {author} {\bibfnamefont {M.}~\bibnamefont {Legner}},
  \bibinfo {author} {\bibfnamefont {A.}~\bibnamefont {R{\"u}egg}},\ and\
  \bibinfo {author} {\bibfnamefont {M.}~\bibnamefont {Sigrist}},\ }\href
  {https://doi.org/10.1103/PhysRevB.95.075116} {\bibfield  {journal} {\bibinfo
  {journal} {Phys. Rev. B}\ }\textbf {\bibinfo {volume} {95}},\ \bibinfo
  {pages} {075116} (\bibinfo {year} {2017})}\BibitemShut {NoStop}%
\bibitem [{\citenamefont {Caio}\ \emph {et~al.}(2019)\citenamefont {Caio},
  \citenamefont {M{\"o}ller}, \citenamefont {Cooper},\ and\ \citenamefont
  {Bhaseen}}]{caio_topological_2019}%
  \BibitemOpen
  \bibfield  {author} {\bibinfo {author} {\bibfnamefont {M.~D.}\ \bibnamefont
  {Caio}}, \bibinfo {author} {\bibfnamefont {G.}~\bibnamefont {M{\"o}ller}},
  \bibinfo {author} {\bibfnamefont {N.~R.}\ \bibnamefont {Cooper}},\ and\
  \bibinfo {author} {\bibfnamefont {M.~J.}\ \bibnamefont {Bhaseen}},\ }\href
  {https://doi.org/10.1038/s41567-018-0390-7} {\bibfield  {journal} {\bibinfo
  {journal} {Nat. Phys.}\ }\textbf {\bibinfo {volume} {15}},\ \bibinfo {pages}
  {257} (\bibinfo {year} {2019})}\BibitemShut {NoStop}%
\bibitem [{\citenamefont {Landau}(1932)}]{landau_theory_1932}%
  \BibitemOpen
  \bibfield  {author} {\bibinfo {author} {\bibfnamefont {L.~D.}\ \bibnamefont
  {Landau}},\ }\href@noop {} {\bibfield  {journal} {\bibinfo  {journal} {Phys.
  Zs. Sowjet}\ }\textbf {\bibinfo {volume} {1}},\ \bibinfo {pages} {88}
  (\bibinfo {year} {1932})}\BibitemShut {NoStop}%
\bibitem [{\citenamefont {Zener}(1932)}]{zener_non-adiabatic_1932}%
  \BibitemOpen
  \bibfield  {author} {\bibinfo {author} {\bibfnamefont {C.}~\bibnamefont
  {Zener}},\ }\href {https://doi.org/10.1098/rspa.1932.0165} {\bibfield
  {journal} {\bibinfo  {journal} {Proc. A}\ }\textbf {\bibinfo {volume}
  {137}},\ \bibinfo {pages} {696} (\bibinfo {year} {1932})}\BibitemShut
  {NoStop}%
\bibitem [{\citenamefont {Halperin}(1982)}]{halperin_quantized_1982}%
  \BibitemOpen
  \bibfield  {author} {\bibinfo {author} {\bibfnamefont {B.~I.}\ \bibnamefont
  {Halperin}},\ }\href {https://doi.org/10.1103/PhysRevB.25.2185} {\bibfield
  {journal} {\bibinfo  {journal} {Phys. Rev. B}\ }\textbf {\bibinfo {volume}
  {25}},\ \bibinfo {pages} {2185} (\bibinfo {year} {1982})}\BibitemShut
  {NoStop}%
\bibitem [{\citenamefont {Hatsugai}(1993)}]{hatsugai_chern_1993}%
  \BibitemOpen
  \bibfield  {author} {\bibinfo {author} {\bibfnamefont {Y.}~\bibnamefont
  {Hatsugai}},\ }\href {https://doi.org/10.1103/PhysRevLett.71.3697} {\bibfield
   {journal} {\bibinfo  {journal} {Phys. Rev. Lett.}\ }\textbf {\bibinfo
  {volume} {71}},\ \bibinfo {pages} {3697} (\bibinfo {year}
  {1993})}\BibitemShut {NoStop}%
\bibitem [{\citenamefont {Lim}\ \emph {et~al.}(2012)\citenamefont {Lim},
  \citenamefont {Fuchs},\ and\ \citenamefont
  {Montambaux}}]{lim_bloch-zener_2012}%
  \BibitemOpen
  \bibfield  {author} {\bibinfo {author} {\bibfnamefont {L.-K.}\ \bibnamefont
  {Lim}}, \bibinfo {author} {\bibfnamefont {J.-N.}\ \bibnamefont {Fuchs}},\
  and\ \bibinfo {author} {\bibfnamefont {G.}~\bibnamefont {Montambaux}},\
  }\href {https://doi.org/10.1103/PhysRevLett.108.175303} {\bibfield  {journal}
  {\bibinfo  {journal} {Phys. Rev. Lett.}\ }\textbf {\bibinfo {volume} {108}},\
  \bibinfo {pages} {175303} (\bibinfo {year} {2012})}\BibitemShut {NoStop}%
\bibitem [{\citenamefont {Tarruell}\ \emph {et~al.}(2012)\citenamefont
  {Tarruell}, \citenamefont {Greif}, \citenamefont {Uehlinger}, \citenamefont
  {Jotzu},\ and\ \citenamefont {Esslinger}}]{tarruell_creating_2012}%
  \BibitemOpen
  \bibfield  {author} {\bibinfo {author} {\bibfnamefont {L.}~\bibnamefont
  {Tarruell}}, \bibinfo {author} {\bibfnamefont {D.}~\bibnamefont {Greif}},
  \bibinfo {author} {\bibfnamefont {T.}~\bibnamefont {Uehlinger}}, \bibinfo
  {author} {\bibfnamefont {G.}~\bibnamefont {Jotzu}},\ and\ \bibinfo {author}
  {\bibfnamefont {T.}~\bibnamefont {Esslinger}},\ }\href
  {https://doi.org/10.1038/nature10871} {\bibfield  {journal} {\bibinfo
  {journal} {Nature}\ }\textbf {\bibinfo {volume} {483}},\ \bibinfo {pages}
  {302} (\bibinfo {year} {2012})}\BibitemShut {NoStop}%
\bibitem [{\citenamefont {Brown}\ \emph {et~al.}(2022)\citenamefont {Brown},
  \citenamefont {Chang}, \citenamefont {Schwarz}, \citenamefont {Leung},
  \citenamefont {Kozii}, \citenamefont {Avdoshkin}, \citenamefont {Moore},\
  and\ \citenamefont {{Stamper-Kurn}}}]{brown_direct_2022}%
  \BibitemOpen
  \bibfield  {author} {\bibinfo {author} {\bibfnamefont {C.~D.}\ \bibnamefont
  {Brown}}, \bibinfo {author} {\bibfnamefont {S.-W.}\ \bibnamefont {Chang}},
  \bibinfo {author} {\bibfnamefont {M.~N.}\ \bibnamefont {Schwarz}}, \bibinfo
  {author} {\bibfnamefont {T.-H.}\ \bibnamefont {Leung}}, \bibinfo {author}
  {\bibfnamefont {V.}~\bibnamefont {Kozii}}, \bibinfo {author} {\bibfnamefont
  {A.}~\bibnamefont {Avdoshkin}}, \bibinfo {author} {\bibfnamefont {J.~E.}\
  \bibnamefont {Moore}},\ and\ \bibinfo {author} {\bibfnamefont
  {D.}~\bibnamefont {{Stamper-Kurn}}},\ }\bibfield  {journal} {\bibinfo
  {journal} {Science}\ }\href {https://doi.org/10.1126/science.abm6442}
  {10.1126/science.abm6442} (\bibinfo {year} {2022})\BibitemShut {NoStop}%
\bibitem [{\citenamefont {Beenakker}(2024)}]{beenakker_chiral_2024}%
  \BibitemOpen
  \bibfield  {author} {\bibinfo {author} {\bibfnamefont {C.~W.~J.}\
  \bibnamefont {Beenakker}},\ }\href
  {https://doi.org/10.1103/PhysRevB.110.165421} {\bibfield  {journal} {\bibinfo
   {journal} {Phys. Rev. B}\ }\textbf {\bibinfo {volume} {110}},\ \bibinfo
  {pages} {165421} (\bibinfo {year} {2024})}\BibitemShut {NoStop}%
\bibitem [{\citenamefont {Zou}\ \emph {et~al.}(2022)\citenamefont {Zou},
  \citenamefont {Fu}, \citenamefont {Wang}, \citenamefont {Hu},\ and\
  \citenamefont {Shen}}]{zou_half-quantized_2022}%
  \BibitemOpen
  \bibfield  {author} {\bibinfo {author} {\bibfnamefont {J.-Y.}\ \bibnamefont
  {Zou}}, \bibinfo {author} {\bibfnamefont {B.}~\bibnamefont {Fu}}, \bibinfo
  {author} {\bibfnamefont {H.-W.}\ \bibnamefont {Wang}}, \bibinfo {author}
  {\bibfnamefont {Z.-A.}\ \bibnamefont {Hu}},\ and\ \bibinfo {author}
  {\bibfnamefont {S.-Q.}\ \bibnamefont {Shen}},\ }\href
  {https://doi.org/10.1103/PhysRevB.105.L201106} {\bibfield  {journal}
  {\bibinfo  {journal} {Phys. Rev. B}\ }\textbf {\bibinfo {volume} {105}},\
  \bibinfo {pages} {L201106} (\bibinfo {year} {2022})}\BibitemShut {NoStop}%
\bibitem [{\citenamefont {Zhuo}\ \emph {et~al.}(2026)\citenamefont {Zhuo},
  \citenamefont {Zhang}, \citenamefont {Zhou}, \citenamefont {Tay},
  \citenamefont {Liu}, \citenamefont {Xi}, \citenamefont {Chen},\ and\
  \citenamefont {Chang}}]{zhuo_half-quantized_2026}%
  \BibitemOpen
  \bibfield  {author} {\bibinfo {author} {\bibfnamefont {D.}~\bibnamefont
  {Zhuo}}, \bibinfo {author} {\bibfnamefont {B.}~\bibnamefont {Zhang}},
  \bibinfo {author} {\bibfnamefont {H.}~\bibnamefont {Zhou}}, \bibinfo {author}
  {\bibfnamefont {H.}~\bibnamefont {Tay}}, \bibinfo {author} {\bibfnamefont
  {X.}~\bibnamefont {Liu}}, \bibinfo {author} {\bibfnamefont {Z.}~\bibnamefont
  {Xi}}, \bibinfo {author} {\bibfnamefont {C.-Z.}\ \bibnamefont {Chen}},\ and\
  \bibinfo {author} {\bibfnamefont {C.-Z.}\ \bibnamefont {Chang}},\ }\href
  {https://doi.org/10.1103/vxcb-rwbl} {\bibfield  {journal} {\bibinfo
  {journal} {Phys. Rev. Lett.}\ }\textbf {\bibinfo {volume} {136}},\ \bibinfo
  {pages} {016601} (\bibinfo {year} {2026})}\BibitemShut {NoStop}%
\bibitem [{\citenamefont {Lepoutre}\ \emph {et~al.}(2019)\citenamefont
  {Lepoutre}, \citenamefont {Schachenmayer}, \citenamefont {Gabardos},
  \citenamefont {Zhu}, \citenamefont {Naylor}, \citenamefont {Mar{\'e}chal},
  \citenamefont {Gorceix}, \citenamefont {Rey}, \citenamefont {Vernac},\ and\
  \citenamefont {{Laburthe-Tolra}}}]{lepoutre_out--equilibrium_2019}%
  \BibitemOpen
  \bibfield  {author} {\bibinfo {author} {\bibfnamefont {S.}~\bibnamefont
  {Lepoutre}}, \bibinfo {author} {\bibfnamefont {J.}~\bibnamefont
  {Schachenmayer}}, \bibinfo {author} {\bibfnamefont {L.}~\bibnamefont
  {Gabardos}}, \bibinfo {author} {\bibfnamefont {B.}~\bibnamefont {Zhu}},
  \bibinfo {author} {\bibfnamefont {B.}~\bibnamefont {Naylor}}, \bibinfo
  {author} {\bibfnamefont {E.}~\bibnamefont {Mar{\'e}chal}}, \bibinfo {author}
  {\bibfnamefont {O.}~\bibnamefont {Gorceix}}, \bibinfo {author} {\bibfnamefont
  {A.~M.}\ \bibnamefont {Rey}}, \bibinfo {author} {\bibfnamefont
  {L.}~\bibnamefont {Vernac}},\ and\ \bibinfo {author} {\bibfnamefont
  {B.}~\bibnamefont {{Laburthe-Tolra}}},\ }\href
  {https://doi.org/10.1038/s41467-019-09699-5} {\bibfield  {journal} {\bibinfo
  {journal} {Nat Commun}\ }\textbf {\bibinfo {volume} {10}},\ \bibinfo {pages}
  {1714} (\bibinfo {year} {2019})}\BibitemShut {NoStop}%
\bibitem [{\citenamefont {Lecomte}\ \emph {et~al.}(2025)\citenamefont
  {Lecomte}, \citenamefont {Journeaux}, \citenamefont {Veschambre},
  \citenamefont {Dalibard},\ and\ \citenamefont
  {Lopes}}]{lecomte_production_2025}%
  \BibitemOpen
  \bibfield  {author} {\bibinfo {author} {\bibfnamefont {M.}~\bibnamefont
  {Lecomte}}, \bibinfo {author} {\bibfnamefont {A.}~\bibnamefont {Journeaux}},
  \bibinfo {author} {\bibfnamefont {J.}~\bibnamefont {Veschambre}}, \bibinfo
  {author} {\bibfnamefont {J.}~\bibnamefont {Dalibard}},\ and\ \bibinfo
  {author} {\bibfnamefont {R.}~\bibnamefont {Lopes}},\ }\href
  {https://doi.org/10.1103/PhysRevLett.134.013402} {\bibfield  {journal}
  {\bibinfo  {journal} {Phys. Rev. Lett.}\ }\textbf {\bibinfo {volume} {134}},\
  \bibinfo {pages} {013402} (\bibinfo {year} {2025})}\BibitemShut {NoStop}%
\bibitem [{\citenamefont {Pasquiou}\ \emph {et~al.}(2010)\citenamefont
  {Pasquiou}, \citenamefont {Bismut}, \citenamefont {Beaufils}, \citenamefont
  {Crubellier}, \citenamefont {Mar{\'e}chal}, \citenamefont {Pedri},
  \citenamefont {Vernac}, \citenamefont {Gorceix},\ and\ \citenamefont
  {{Laburthe-Tolra}}}]{pasquiou_control_2010}%
  \BibitemOpen
  \bibfield  {author} {\bibinfo {author} {\bibfnamefont {B.}~\bibnamefont
  {Pasquiou}}, \bibinfo {author} {\bibfnamefont {G.}~\bibnamefont {Bismut}},
  \bibinfo {author} {\bibfnamefont {Q.}~\bibnamefont {Beaufils}}, \bibinfo
  {author} {\bibfnamefont {A.}~\bibnamefont {Crubellier}}, \bibinfo {author}
  {\bibfnamefont {E.}~\bibnamefont {Mar{\'e}chal}}, \bibinfo {author}
  {\bibfnamefont {P.}~\bibnamefont {Pedri}}, \bibinfo {author} {\bibfnamefont
  {L.}~\bibnamefont {Vernac}}, \bibinfo {author} {\bibfnamefont
  {O.}~\bibnamefont {Gorceix}},\ and\ \bibinfo {author} {\bibfnamefont
  {B.}~\bibnamefont {{Laburthe-Tolra}}},\ }\href
  {https://doi.org/10.1103/PhysRevA.81.042716} {\bibfield  {journal} {\bibinfo
  {journal} {Phys. Rev. A}\ }\textbf {\bibinfo {volume} {81}},\ \bibinfo
  {pages} {042716} (\bibinfo {year} {2010})}\BibitemShut {NoStop}%
\bibitem [{\citenamefont {Barral}\ \emph {et~al.}(2024)\citenamefont {Barral},
  \citenamefont {Cantara}, \citenamefont {Du}, \citenamefont {Lunden},
  \citenamefont {{de Hond}}, \citenamefont {Jamison},\ and\ \citenamefont
  {Ketterle}}]{barral_suppressing_2024}%
  \BibitemOpen
  \bibfield  {author} {\bibinfo {author} {\bibfnamefont {P.}~\bibnamefont
  {Barral}}, \bibinfo {author} {\bibfnamefont {M.}~\bibnamefont {Cantara}},
  \bibinfo {author} {\bibfnamefont {L.}~\bibnamefont {Du}}, \bibinfo {author}
  {\bibfnamefont {W.}~\bibnamefont {Lunden}}, \bibinfo {author} {\bibfnamefont
  {J.}~\bibnamefont {{de Hond}}}, \bibinfo {author} {\bibfnamefont {A.~O.}\
  \bibnamefont {Jamison}},\ and\ \bibinfo {author} {\bibfnamefont
  {W.}~\bibnamefont {Ketterle}},\ }\href
  {https://doi.org/10.1038/s41467-024-47260-1} {\bibfield  {journal} {\bibinfo
  {journal} {Nat Commun}\ }\textbf {\bibinfo {volume} {15}},\ \bibinfo {pages}
  {3566} (\bibinfo {year} {2024})}\BibitemShut {NoStop}%
\bibitem [{\citenamefont {Gross}\ and\ \citenamefont
  {Neveu}(1974)}]{gross_dynamical_1974}%
  \BibitemOpen
  \bibfield  {author} {\bibinfo {author} {\bibfnamefont {D.~J.}\ \bibnamefont
  {Gross}}\ and\ \bibinfo {author} {\bibfnamefont {A.}~\bibnamefont {Neveu}},\
  }\href {https://doi.org/10.1103/PhysRevD.10.3235} {\bibfield  {journal}
  {\bibinfo  {journal} {Phys. Rev. D}\ }\textbf {\bibinfo {volume} {10}},\
  \bibinfo {pages} {3235} (\bibinfo {year} {1974})}\BibitemShut {NoStop}%
\bibitem [{\citenamefont {Tabatabaei}\ \emph {et~al.}(2022)\citenamefont
  {Tabatabaei}, \citenamefont {Negari}, \citenamefont {Maciejko},\ and\
  \citenamefont {Vaezi}}]{tabatabaei_chiral_2022}%
  \BibitemOpen
  \bibfield  {author} {\bibinfo {author} {\bibfnamefont {S.~M.}\ \bibnamefont
  {Tabatabaei}}, \bibinfo {author} {\bibfnamefont {A.-R.}\ \bibnamefont
  {Negari}}, \bibinfo {author} {\bibfnamefont {J.}~\bibnamefont {Maciejko}},\
  and\ \bibinfo {author} {\bibfnamefont {A.}~\bibnamefont {Vaezi}},\ }\href
  {https://doi.org/10.1103/PhysRevLett.128.225701} {\bibfield  {journal}
  {\bibinfo  {journal} {Phys. Rev. Lett.}\ }\textbf {\bibinfo {volume} {128}},\
  \bibinfo {pages} {225701} (\bibinfo {year} {2022})}\BibitemShut {NoStop}%
\bibitem [{\citenamefont {Gao}\ \emph {et~al.}(2025)\citenamefont {Gao},
  \citenamefont {Wang},\ and\ \citenamefont {Lee}}]{gao_interacting_2025}%
  \BibitemOpen
  \bibfield  {author} {\bibinfo {author} {\bibfnamefont {Z.-Q.}\ \bibnamefont
  {Gao}}, \bibinfo {author} {\bibfnamefont {T.}~\bibnamefont {Wang}},\ and\
  \bibinfo {author} {\bibfnamefont {D.-H.}\ \bibnamefont {Lee}},\ }\href@noop
  {} {\bibfield  {journal} {\bibinfo  {journal} {arXiv:2504.15338}\ } (\bibinfo
  {year} {2025})},\ \Eprint {https://arxiv.org/abs/2504.15338}
  {arXiv:2504.15338} \BibitemShut {NoStop}%
\bibitem [{\citenamefont {Kitagawa}\ and\ \citenamefont
  {Ueda}(1993)}]{kitagawa_squeezed_1993}%
  \BibitemOpen
  \bibfield  {author} {\bibinfo {author} {\bibfnamefont {M.}~\bibnamefont
  {Kitagawa}}\ and\ \bibinfo {author} {\bibfnamefont {M.}~\bibnamefont
  {Ueda}},\ }\href {https://doi.org/10.1103/PhysRevA.47.5138} {\bibfield
  {journal} {\bibinfo  {journal} {Phys. Rev. A}\ }\textbf {\bibinfo {volume}
  {47}},\ \bibinfo {pages} {5138} (\bibinfo {year} {1993})}\BibitemShut
  {NoStop}%
\bibitem [{\citenamefont {Peskin}(2018)}]{peskin_introduction_2018}%
  \BibitemOpen
  \bibfield  {author} {\bibinfo {author} {\bibfnamefont {M.~E.}\ \bibnamefont
  {Peskin}},\ }\href {https://doi.org/10.1201/9780429503559} {\emph {\bibinfo
  {title} {An {{Introduction To Quantum Field Theory}}}}}\ (\bibinfo
  {publisher} {CRC Press},\ \bibinfo {address} {Boca Raton},\ \bibinfo {year}
  {2018})\BibitemShut {NoStop}%
\bibitem [{\citenamefont {Sachdev}(2011)}]{sachdev_quantum_2011}%
  \BibitemOpen
  \bibfield  {author} {\bibinfo {author} {\bibfnamefont {S.}~\bibnamefont
  {Sachdev}},\ }\href {https://doi.org/10.1017/CBO9780511973765} {\emph
  {\bibinfo {title} {Quantum {{Phase Transitions}}}}},\ \bibinfo {edition}
  {2nd}\ ed.\ (\bibinfo  {publisher} {Cambridge University Press},\ \bibinfo
  {year} {2011})\BibitemShut {NoStop}%
\end{thebibliography}
\end{document}